\documentclass[prd,preprint,superscriptaddress,showpacs,nofootinbib,%
tightenlines ]{revtex4}
\usepackage{graphics}
\usepackage{epsfig}
\usepackage{amsmath,amssymb}
\usepackage{slashed}

\newcommand{\WI}{Ward identities }
\newcommand{\psl}{\slashed{p}}
\newcommand{\ti}{\tilde}
\newcommand{\mt}{\tilde{m}}
\newcommand{\st}{\widetilde{\Sigma}}
\newcommand{\RT}{\tilde{R}}
\newcommand{\IT}{\tilde{I}}
\newcommand{\logM}{\ln\frac{M}{\mu}}
\newcommand{\logMsq}{\ln^2\frac{M}{\mu}}

\begin{document}
\preprint{MKPH-T-07-10}
\title{Infrared renormalization of two-loop integrals and the chiral expansion of the nucleon mass}
\author{Matthias R.~Schindler}
\altaffiliation[Current address: ]{Department of Physics \& Astronomy, Ohio University, Athens, OH 45701, USA}
\affiliation{Institut f\"ur Kernphysik, Johannes
Gutenberg-Universit\"at, 55099 Mainz, Germany}
\author{Dalibor Djukanovic}
\affiliation{Institut f\"ur Kernphysik, Johannes
Gutenberg-Universit\"at, 55099 Mainz, Germany}
\author{Jambul Gegelia}
\affiliation{Institut f\"ur Kernphysik, Johannes
Gutenberg-Universit\"at, 55099 Mainz, Germany} \affiliation{High
Energy Physics Institute, Tbilisi State University, Tbilisi,
Georgia}
\author{Stefan Scherer}
\affiliation{Institut f\"ur Kernphysik, Johannes
Gutenberg-Universit\"at, 55099 Mainz, Germany}
\begin{abstract}
   We describe details of the renormalization of two-loop
integrals relevant to the calculation of the nucleon mass in the
framework of manifestly Lorentz-invariant chiral perturbation
theory using infrared renormalization.
   It is shown that the renormalization can be performed while
preserving all relevant symmetries, in particular chiral symmetry,
and that renormalized diagrams respect the standard power counting
rules.
   As an application we calculate the chiral expansion of the nucleon
mass to order ${\cal O}(q^6)$.
\end{abstract}
\date{July 30, 2007}
\maketitle

\section{Introduction}
   Chiral perturbation theory (ChPT)
\cite{Weinberg:1978kz,Gasser:1983yg,Gasser:1984gg} is the
effective field theory of the strong interactions at low energies
(for an introduction see, e.g.,
\cite{Scherer:2002tk,Scherer:2005ri,Bernard:2007zu}).
   It relies on a perturbative expansion in terms of small
parameters $q/\Lambda$, where $q$ denotes a quantity like the pion
mass or external momenta that are small relative to the scale
$\Lambda$, which for ChPT is expected to be of the size
$1\,\mbox{GeV}$.
   One of the essential ingredients of ChPT is a consistent power
counting, which assigns a chiral order $D$ to each Feynman diagram
for the process in question and which predicts that diagrams of
higher orders are suppressed.
   Assuming the coefficients of the perturbative expansion to be
of natural size one would expect contributions at order $D+1$ to
be suppressed by a factor $q/\Lambda$ compared to contributions at
order $D$.
   For $q$ of the order of the pion mass and
$\Lambda\approx1\,\mbox{GeV}$, this corresponds to a correction of
about $20\%$.
   In the mesonic sector of ChPT this rough estimate seems accurate,
however, the situation is less clear for the baryonic sector.
   While for example the chiral expansion of the nucleon mass
shows a good convergence behavior, the nucleon axial coupling
$g_A$ receives large contributions from higher-order terms
\cite{Kambor:1998pi}.
   Further examples include the electromagnetic form factors of
the nucleon (see, e.g., \cite{Kubis:2000zd,Schindler:2005ke}),
which only describe the data for very low values of momentum
transfer.
   For some of these quantities higher-order contributions clearly
play an important role.
   The description of the nucleon form factors can be improved by the
inclusion of vector mesons as explicit degrees of freedom, which
corresponds to the resummation of higher-order contributions
\cite{Kubis:2000zd,Schindler:2005ke}.

   The convergence properties of baryon chiral perturbation
theory (BChPT) are also of great importance for lattice QCD.
  Due to numerical costs, present lattice calculations still
require pion masses larger than the physical one, and results
obtained on the lattice have to be extrapolated to the physical
point.
   ChPT as an expansion in the pion mass is the appropriate tool to
perform such extrapolations, which again poses the question for
which values of small parameters the ChPT expansion gives reliable
predictions.

   There are several renormalization schemes for
manifestly Lorentz-invariant BChPT at the one-loop level that
result in a consistent power counting while preserving all
relevant symmetries
\cite{Ellis:1997kc,Becher:1999he,Gegelia:1999gf,Gegelia:1999qt,Goity:2001ny,Fuchs:2003qc}.
   The most commonly used of these is the infrared (IR) regularization
of Ref.~\cite{Becher:1999he}.
   All these renormalization schemes have in
common that there is a relation between the chiral order and the
loop expansion, so that the investigation of higher chiral orders
requires the evaluation of multi-loop diagrams.
   In Ref.~\cite{Schindler:2003xv} a reformulated version of the IR
regularization has been introduced that is also applicable to
multi-loop diagrams \cite{Schindler:2003je}.
   Reference~\cite{Lehmann:2001xm} contains a different generalization of IR
regularization to two-loop diagrams.

   In this paper we describe the renormalization
procedure for two-loop integrals in manifestly Lorentz-invariant
BChPT within the reformulated version of infrared regularization
of Ref.~\cite{Schindler:2003xv}.
   The results of a calculation of the nucleon mass up to and
including order ${\cal O}(q^6)$ have been reported in
Ref.~\cite{Schindler:2006ha}.
   To the best of our knowledge, this is the first complete
two-loop BChPT calculation in a manifestly Lorentz-invariant
framework.
   Here, we describe the details of the calculation.
   In particular we show that the renormalization procedure preserves
all relevant symmetries and that renormalized two-loop diagrams
obey the standard power counting rules.
   A calculation of the nucleon mass to order ${\cal
O}(q^5)$ was performed in Ref.~\cite{McGovern:1998tm} in the
framework of HBChPT (see, e.g.,
\cite{Jenkins:1990jv,Bernard:1992qa}), and
Ref.~\cite{Bernard:2006te} contains the leading non-analytic
contributions to the axial-vector coupling $g_A$ at two-loop order
obtained from renormalization group techniques.

   This paper is organized as follows.
   In Section~\ref{sec:IROneLoop} we review the main features of
infrared renormalization at the one-loop level that are essential
for the following.
   Section~\ref{sec:GenTwoLoop} contains a brief overview over the
general aspects of the renormalization of two-loop diagrams.
   The infrared renormalization of products of one-loop integrals
is described in Sec.~\ref{sec:OneLoopProducts}, while the
discussion of genuine two-loop integrals follows in
Sec.~\ref{Sec:TwoLoop}.
   An application of these methods can be found in
Sec.~\ref{sec:NucMass}, which is followed by a summary.
   Explicit expressions for the appearing integrals can be found
in the appendix.

\section{Infrared regularization of one-loop
integrals}\label{sec:IROneLoop}

   The method of infrared regularization \cite{Becher:1999he} was
developed as a manifestly Lorentz-invariant renormalization scheme
preserving all relevant symmetries.
   It results in diagrams obeying the standard power counting (see Sec.~\ref{sec:NucMass}).
   Infrared regularization is based on dimensional regularization and the analytic properties
of loop integrals, and in its original formulation is applicable
to one-loop integrals containing pion and nucleon propagators in
the one-nucleon sector of ChPT.
   In Ref.~\cite{Schindler:2003xv} a different formulation of
infrared regularization was presented which reproduces the results
of the original formulation up to arbitrary order.
   The advantage of the new formulation is that it can also be applied
to multi-loop diagrams and diagrams containing additional degrees
of freedom
\cite{Fuchs:2003sh,Schindler:2003xv,Schindler:2003je,Schindler:2005ke,Schindler:2006it}.
   Since IR regularization can, in fact, be viewed as a renormalization
scheme we also refer to it as infrared renormalization.
   We briefly describe those features of the renormalization of
one-loop integrals which are important for the renormalization of
two-loop integrals.

   Denote a general one-loop integral containing pion and nucleon
propagators by
\begin{equation}\label{IR:GenInt}
  H_{\pi\cdots N\cdots}(q_1,\ldots,p_1,\ldots)=i \int
\frac{d^nk}{(2\pi)^n} \ \frac{1}{a_1\cdots a_m \ b_1\cdots b_l},
\end{equation}
   where $a_i=(k+q_i)^2-M^2+i 0^+$ and $b_j=(k+p_j)^2-m^2+i 0^+$
are related to pion and nucleon propagators, respectively, and
$n=4+2\epsilon$ is the number of space-time dimensions.
   Infrared renormalization consists of splitting the integral
into an infrared singular part $I_{\pi\cdots N\cdots}$ and an
infrared regular part $R_{\pi\cdots N\cdots}$,
\begin{equation}\label{IR:IpiN+RpiN}
    H_{\pi\cdots N\cdots}=I_{\pi\cdots N\cdots}+R_{\pi\cdots
    N\cdots}\,,
\end{equation}
or for short
\begin{equation}\label{IR:I+R}
    H=I+R\,.
\end{equation}
   The advantage of splitting the original integral into two parts is
that the infrared singular part $I_{\pi\cdots N\cdots}$ satisfies
the power counting, while $R_{\pi\cdots N\cdots}$ contains terms
that violate the power counting.
   In addition, the infrared singular and infrared regular parts differ in their
analytic properties.
   For noninteger $n$ the expansion of $I_{\pi\cdots N\cdots}$ in small quantities
results in only noninteger powers of these variables, while
$R_{\pi\cdots N\cdots}$ only contains analytic contributions.
   In the formulation of Ref.~\cite{Schindler:2003xv} one obtains
the infrared regular part $R_{\pi\cdots N\cdots}$ by reducing
$H_{\pi\cdots N\cdots}$ to an integral over Schwinger or Feynman
parameters, expanding the resulting expression in small quantities
such as pion masses or small momenta, and interchanging summation
and integration.

   As an example consider the integral
\begin{equation}\label{IR:ExInt}
H_{{\pi}N}(0,-p)=i\int\frac{d^nk}{(2\pi)^n}\frac{1}{[k^2-M^2+i0^+]
[(k-p)^2-m^2+i 0^+]}.
\end{equation}
   To apply the reformulated version of IR renormalization we
combine the two propagators using
$$ \frac{1}{a b} = \int_0^1 \frac{dz}{[(1-z)a+z b]^2}\,, $$
and perform the integration over the loop momentum $k$, resulting
in
\begin{equation}\label{IR:intpar}
H_{{\pi}N}(0,-p)=-\frac{1}{(4\pi)^{n/2}} \ \Gamma(2-n/2) \int_0^1
dz \ \left[ C(z)\right]^{(n/2)-2},
\end{equation}
where $C(z)=m^2 z^2-(p^2-m^2) (1-z) z+M^2 (1-z)-i 0^+$.
   Next, we expand $\left[ C(z)\right]^{(n/2)-2}$ in $p^2-m^2$ and $M^2$ and interchange
summation and integration.
   This generates the chiral expansion of the infrared regular part
$R$, which is given by
\begin{equation}\label{IR:Rrepr}
R=-\frac{m^{n-4} \Gamma(2-n/2)}{(4\pi)^{n/2} (n-3)}
\left[1-\frac{p^2-m^2}{2 m^2}+\frac{(n-6)
\left(p^2-m^2\right)^2}{4 m^4 (n-5)}+\frac{(n-3) M^2 }{2 m^2 (n-5)
}+\cdots\right],
\end{equation}
   and which coincides with the expansion of $R$
given in Ref.~\cite{Becher:1999he}.

   Symmetries introduce relations among various Green functions of
the theory, called Ward-Fradkin-Takahashi identities (Ward
identities for short)
\cite{Ward:1950xp,Fradkin:1955jr,Takahashi:1957xn}.
   Expressions containing the integrals $H$ satisfy the Ward
identities,\footnote{In the following we use the phrase that
integrals satisfy the Ward identities, by which we mean that
expressions containing these integrals satisfy the Ward
identities.} since they are derived from an invariant Lagrangian
and dimensional regularization does not violate the symmetries.
   Since $I$, for noninteger $n$, only contains nonanalytic terms, while $R$ consists
of analytic contributions only, each part has to satisfy the Ward
identities separately in order for the sum $H=I+R$ not to violate
any symmetry.

   Both the infrared regular and the infrared singular parts contain
additional divergences not present in the original integral $H$.
   Since these additional divergences do not appear in $H$,
they have to cancel in the sum of $I+R=H$.
   This means that
\begin{equation}\label{IR:AddDiv}
   \frac{R^{add}}{\epsilon}=-\frac{I^{add}}{\epsilon}\,.
\end{equation}

   The $\epsilon$ expansion of $H$ is given by
\begin{eqnarray}\label{IR:Hepsilon}
   H &=& \frac{H^{UV}}{\epsilon}+H^{(0)}+{\cal O}(\epsilon)
\nonumber\\
   &=& \frac{H^{UV}}{\epsilon}+ \frac{I^{add}}{\epsilon} +\frac{R^{add}}{\epsilon} + \IT
   +\bar{R},
\end{eqnarray}
   where $H^{UV}/\epsilon$ denotes the ultraviolet divergence of
$H$, $H^{(0)}$ refers to the terms independent of $\epsilon$, and
we have explicitly shown the additional divergences in the second
line.
   In BChPT the renormalization can be performed in a two-step
process.
   First, all divergences are absorbed, and then additional finite
terms are subtracted.
   In the standard approach the divergences are absorbed using the
$\widetilde{\rm MS}$ scheme.
   In this scheme one subtracts the quantity
$$
\frac{1}{32\pi^2}\left[ \frac{1}{\epsilon} - \ln(4\pi)
+\gamma_E-1\right],
$$
where $\gamma_E=-\Gamma'(1)$, and sets the appearing t'Hooft
parameter \cite{'tHooft:1973mm} $\mu=m$, where $m$ is the nucleon
mass in the chiral limit.
   Here, in order to simplify the calculation, we apply minimal
subtraction (MS) with a t'Hooft parameter $\tilde\mu$, absorbing
only terms proportional to $\epsilon^{-1}$, and then set
$\tilde\mu=\frac{m}{(4\pi)^{1/2}}\, e^{\frac{\gamma_E-1}{2}}$.
   This is completely equivalent to the standard approach.
   The infrared renormalized expression $H^r$ of the integral $H$ is defined as
its finite infrared singular term,
\begin{equation}\label{IR:Itilde}
   H^r =  \IT,
\end{equation}
which satisfies the power counting since all terms violating it
are contained in $R$.
   One of the fundamental properties used in the construction of
the effective Lagrangian is the invariance under symmetries of the
underlying theory.
   It is therefore of utmost importance that these symmetries are
not violated at any step in the calculations.
   We now show that the definition of the renormalized integral $H^r$
of Eq.~(\ref{IR:Itilde}) satisfies this requirement
\cite{Becher:1999he}.
   The original integral $H$ is obtained from a chirally
symmetric Lagrangian using dimensional regularization, which
preserves all symmetries.
   Therefore expressions containing $H$ satisfy the Ward
identities; and in particular their $\epsilon$ expansions satisfy
the \WI order by order.
   As explained above, $R$ satisfies the \WI separately from
$I$.
   This also means that the \WI are satisfied order by order in
the $\epsilon$ expansion of $R$ and $I$, respectively.
   Therefore the identification of the renormalized integral $H^r$
as $H^r=\IT$ does not violate any symmetry constraints.
   Since the sum of additional divergences cancels, the
term which is subtracted from $H$ is given by
\begin{equation}\label{IR:Rtilde}
    \RT=\frac{H^{UV}}{\epsilon}+ \bar{R}.
\end{equation}
   With Eq.~(\ref{IR:AddDiv}) and the definition of Eq.~(\ref{IR:Rtilde})
we can write
\begin{equation}\label{IR:HItildeRtilde}
    H= \IT +\RT.
\end{equation}

   Within the framework of dimensional regularization, the
dimensional counting analysis of Ref.~\cite{Gegelia:1994zz}
provides a method to obtain expansions of loop integrals in small
parameters.
   This method is described in detail in Appendix~\ref{App:DimCount}.
   Here we show how the infrared regular and infrared singular
parts of the integral $H$ are related to the different terms
obtained from this method.
   Using dimensional counting, $H$ is written as \footnote{Note that the notation used here differs from the one found in the Appendix to avoid confusion with terms in the $\epsilon$ expansion of $H$.}
\begin{equation}\label{Ren:F1+F2}
    H=G_1+G_2.
\end{equation}
   For $G_1$ we simply expand the integrand in $M$ and interchange
summation and integration.
   $G_2$ is obtained by rescaling the integration variable $k\mapsto\frac{M}{m}\,k$ and
then expanding the integrand with subsequent interchange of
summation and integration.
   For $p^2=m^2$ the method of obtaining $G_1$ is the same as the one used to
determine the expansion of the infrared-regular part $R$.
   It follows that
\begin{equation}\label{Ren:F1R}
  G_1 = \sum_n R_n = R,
\end{equation}
while $G_2$ gives the chiral expansion of the infrared singular
term $I$,
\begin{equation}\label{Ren:F2I}
  G_2 =\sum_n I_n,
\end{equation}
where $R_n$ and $I_n$ are the terms in the chiral expansion of the
infrared regular and infrared singular parts, respectively.
   It should be noted that the expansion of $I$ does not always
converge in the entire low-energy region \cite{Becher:1999he}.
   For the integrals considered in the calculation of the nucleon
mass, however, the expansion of $I$ converges.
   The identification of $G_1$ and $G_2$ with the infrared regular
and infrared singular parts, respectively, is used below to show
that the renormalization process in the two-loop sector does not
violate the considered symmetries.

\section{General features of the renormalization of two-loop
integrals}\label{sec:GenTwoLoop}

 We give a brief description of the general renormalization procedure
for two-loop integrals before presenting details of the IR
renormalization.
   The discussion follows Ref.~\cite{Collins:1984xc}.

\begin{figure}
\begin{center}
\epsfig{file=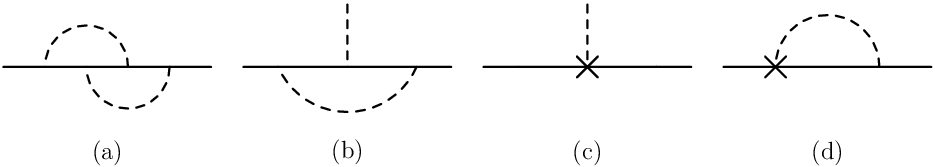,width=0.9\textwidth}
\end{center}
\caption{Two-loop diagram with corresponding subdiagram and
counterterm diagram.\label{Ren:ExPic}}
\end{figure}

   At the two-loop level integrals not only contain overall
UV divergences, but can also contain subdivergences for the case
where one integration momentum is fixed while the other one goes
to infinity.
   As an example consider the two-loop diagram of
Fig.~\ref{Ren:ExPic}~(a).
   It contains one-loop subdiagrams, shown in
Fig.~\ref{Ren:ExPic}~(b).
   The renormalization of subdiagrams requires vertices
as shown in Fig.~\ref{Ren:ExPic}~(c), which are of order
$\hbar$.\footnote{Here, the power of $\hbar$ denotes the order in
the loop expansion.}
   At order $\hbar^2$ these vertices appear in so-called counterterm
diagrams as the one shown in Fig.~\ref{Ren:ExPic}~(d).
   When the sum of the original diagram and the one-loop counterterm diagrams,
Fig.~\ref{Ren:ExPic}~(a) and twice the contribution from
Fig.~\ref{Ren:ExPic}~(d), respectively, is considered, the
remaining divergence is local and can be absorbed by counterterms.
   In order to renormalize a two-loop diagram one has to
take into account all corresponding one-loop counterterm diagrams.

   In our calculation we encounter two general types of two-loop integrals.
   The first type can be directly written as the product of two
one-loop integrals, while this decomposition is not possible for
the second type.

\section{Infrared renormalization of products of one-loop
integrals}\label{sec:OneLoopProducts}

   Consider the product of two one-loop integrals,
\begin{equation}
   H=H_a H_b.
\end{equation}
   $H$ is a two-loop integral and the result of a dimensional
counting analysis reads (see App.~\ref{App:DimCountTL})
\begin{align}
   H & = H^{(0,0)}+H^{(1,0)}+H^{(0,1)}+H^{(1,1)}  \notag\\
& = F_1+F_2+F_3+F_4,
\end{align}
where $F_1$, $F_2+F_3$, and $F_4$ satisfy the Ward identities
separately due to different analytic structures, i.e. different
overall powers of $M$ in $n$ dimensions.
   Using Eq.~(\ref{IR:I+R}), $H$ can also be expressed as
\begin{equation}
   H=I_a I_b + I_a R_b + R_a I_b + R_a R_b,
\end{equation}
where again $I_a I_b$, $I_a R_b + R_a I_b$, and $R_a R_b$ satisfy
the Ward identities individually.

   To renormalize the integral $H$ we need to
add the contributions of (renormalized) counterterm integrals.
   The vertex used in the counterterm integral is determined by standard IR
renormalization of a one-loop subintegral.
   In a one-loop calculation we do not have to consider terms proportional
to $\epsilon$ for the subtraction terms since, at the end of the
calculation, the limit $\epsilon\rightarrow 0$ is taken.
   At the two-loop level, however, the subtraction terms are multiplied
with terms proportional to $\epsilon^{-1}$ from the second loop
integration.
   Therefore the choice whether or not to include the terms proportional to $\epsilon$
in one-loop subtraction terms results in different finite
contributions in the two-loop integrals.
   In addition to the UV divergences and the terms proportional to $\epsilon^0$
we choose the subtraction terms for one-loop integrals to contain
all positive powers of $\epsilon$,\footnote{In a calculation at
two-loop order it is sufficient to include terms proportional to
$\epsilon$. Higher powers of $\epsilon$ are required for the
generalization to multi-loop diagrams.}
\begin{equation}\label{Ren:RTilde}
    \RT=\frac{H^{UV}}{\epsilon}+\RT^{(0)}+\epsilon\RT^{(1)}+\cdots.
\end{equation}
   This choice is crucial for the preservation of the relevant symmetries as
is discussed in the following.
   $H$ contains two subintegrals, $H_a$ and $H_b$.
   The expressions for the unrenormalized counterterm integrals then read
\begin{equation}
   -\RT_a H_b - \RT_b H_a,
\end{equation}
which themselves need to be renormalized applying IR
renormalization.
   The $H_i$ $(i=a,b)$ are one-loop integrals from which we would subtract
the term $\RT_i$ in a one-loop calculation, excluding the
additional divergences.
   However, the term $\RT_j$ multiplying $H_i$ contains terms with positive
powers of $\epsilon$, so that in the product of $\RT_j$ and $R_i$
we get finite terms from the additional divergences in $R_i$ (see
Eq.~(\ref{IR:AddDiv})).
   These would not be removed if we chose the subtraction term to
be $\RT_j \RT_i$.
   Instead we define the subtraction term for the product $\RT_j
H_i$ to be
\begin{equation}\label{Ren:SubSubTerms}
    -\RT_j R_i + \frac{H_j^{UV}R_i^{add}}{\epsilon^2} +\RT_j^0 \frac{R_i^{add}}{\epsilon},
\end{equation}
i.e. we subtract all finite terms stemming from the additional
divergences in $R_i$ but do not subtract the additional
divergences themselves.
   This is analogous to the one-loop sector, where we do not
subtract the additional divergences in the infrared regular part
either (see Eqs.~(\ref{IR:Hepsilon}) and (\ref{IR:Rtilde})).

   We now show that this renormalization procedure for the
counterterm integrals does not violate the Ward identities.
   We know that the subtraction terms $S$ for one-loop integrals do not
violate the Ward identities and result in a modification of the
coupling constants and fields in the Lagrangian.
   The counterterm integrals are then calculated with the help of this new
Lagrangian which means that the term
\begin{equation}\label{Ren:SubSubtermsS}
    -S \, H_i
\end{equation}
also respects all Ward identities.
   $H_i$ is a one-loop integral and Eq.~(\ref{Ren:SubSubtermsS}) can be written as
\begin{equation}\label{Ren:SubSubtermsSI+R}
    -S\, I_i - S\, R_i
\end{equation}
where  $-S\, I_i$ and $-S\, R_i$ satisfy the Ward identities
separately.
   In particular, the Ward identities are satisfied term by term in an expansion in
$\epsilon$ for $S\, I_i$ and $S\, R_i$, respectively.
   The expansion for $S\,I_i$ is given by
\begin{equation}\label{Ren:SIExp}
    S\,I_i = \left(\frac{S^{div}}{\epsilon}+S^{fin}\right)
\left(\frac{I_i^{add}}{\epsilon}+I_i^{fin}\right)
   = \frac{S^{div}I_i^{add}}{\epsilon^2}
+\frac{1}{\epsilon}\left[S^{div}I_i^{fin}+S^{fin}I_i^{add}\right]+\cdots.
\end{equation}
   Suppose we choose the finite part of the counterterm to vanish,\footnote{
In baryonic ChPT this would result in terms violating the power
counting. So far we are only concerned with the symmetries of the
theory, which are conserved for $S^{fin}=0$. The issue of power
counting is addressed below.}
$$S^{fin}=0.$$
   In this case we can see that the term proportional to $\epsilon^{-1}$
is given by
\begin{equation}\label{Ren:SdivIfin}
    \frac{1}{\epsilon}\,S^{div}\,I_i^{fin}.
\end{equation}
   It has to satisfy the Ward identities since for this choice of $S$ it
is the only term proportional to $\epsilon^{-1}$ in the $\epsilon$
expansion of $S\,I_i$.
   By changing the renormalization scheme to also include finite
terms in the subtraction terms, the product in
Eq.~(\ref{Ren:SdivIfin}) does not change, but we obtain the more
general expression of Eq.~(\ref{Ren:SIExp}).
   Considering the term proportional to $\epsilon^{-1}$ and keeping in mind that
Eq.~(\ref{Ren:SdivIfin}) respects the Ward identities we now see
that
\begin{equation}\label{Ren:SfinIdiv}
    \frac{1}{\epsilon}\,S^{fin}\,I_i^{add}
\end{equation}
satisfies the Ward identities separately.
   Since the additional divergences have to cancel in the sum of $I_i$ and $R_i$
it follows that $I_i^{add}=-R_i^{add}$ and
\begin{equation}\label{Ren:SfinRdiv}
    -\frac{1}{\epsilon}\,S^{fin}\,R_i^{add}
\end{equation}
does not violate any symmetry constraints.
   Using the fact that $S\, R_i$ respects all symmetries and choosing
the subtraction term $S$ to be $\RT_j$ (which only contains UV
divergences),
$$S=\RT_j,\quad S^{div}=H_j^{UV},\quad S^{fin}=\RT_j^{(0)}, $$
it follows that
\begin{equation}
    -\frac{H_j^{UV}R_i^{add}}{\epsilon^2}-\RT_j^{(0)} \frac{R_i^{add}}{\epsilon}
\end{equation}
satisfies the Ward identities and therefore also our prescription
for the subtraction terms of the counterterm diagrams of
Eq.~(\ref{Ren:SubSubTerms}) satisfies the Ward identities.

   Using the above method the sum of the original expression and the renormalized
counterterm integrals gives
\begin{eqnarray}
   \lefteqn{H_a H_b -\RT_a H_b + \RT_a R_b -\frac{H_a^{UV}R_b^{add}}{\epsilon^2} -\RT_a^{(0)} \frac{R_b^{add}}{\epsilon}
-\RT_b H_a + \RT_b R_a -\frac{R_a^{add}H_b^{UV}}{\epsilon^2}  } \nonumber\\
   && - \RT_b^{(0)} \frac{R_a^{add}}{\epsilon} \nonumber\\
   &=& I_a I_b + I_a R_b +I_b R_a + R_a R_b -\RT_a I_b -\frac{H_a^{UV}R_b^{add}}{\epsilon^2}
- \RT_a^{(0)} \frac{R_b^{add}}{\epsilon}-\RT_b I_a \nonumber\\
   && -\frac{R_a^{add}H_b^{UV}}{\epsilon^2} - \RT_b^{(0)} \frac{R_a^{add}}{\epsilon} \nonumber \\
   &=& I_a I_b + I_a (R_b-\RT_b) +I_b (R_a-\RT_a)
-\frac{H_a^{UV}R_b^{add}+R_a^{add}H_b^{UV}}{\epsilon^2} -\RT_a^{(0)} \frac{R_b^{add}}{\epsilon}\nonumber\\
   &&  - \RT_b^{(0)}
\frac{R_a^{add}}{\epsilon}+R_aR_b\,.
\end{eqnarray}
   The difference between $R_i$ and $\RT_i$ is only given by the
additional divergences $R_i^{add}/\epsilon$, resulting in
\begin{eqnarray}
   \lefteqn{ \left(\IT_a+\frac{I_a^{add}}{\epsilon}\right)\left(\IT_b+\frac{I_b^{add}}{\epsilon}\right)
+\left(\IT_a+\frac{I_a^{add}}{\epsilon}\right)\frac{R_b^{add}}{\epsilon}
+\left(\IT_b+\frac{I_b^{add}}{\epsilon}\right)\frac{R_a^{add}}{\epsilon}
}  \nonumber\\
   &&  -\frac{H_a^{UV}R_b^{add}+R_a^{add}H_b^{UV}}{\epsilon^2}-\RT_a^{(0)} \frac{R_b^{add}}{\epsilon}
- \RT_b^{(0)} \frac{R_a^{add}}{\epsilon}+R_aR_b\,.\hspace{3cm}
\end{eqnarray}
   Using $I_i^{add}=-R_i^{add}$ we obtain
\begin{equation}
    \IT_a \IT_b -\frac{I_a^{add}I_b^{add}}{\epsilon^2}-\frac{H_a^{UV}R_b^{add}+R_a^{add}H_b^{UV}}{\epsilon^2}
-\RT_a^{(0)} \frac{R_b^{add}}{\epsilon} - \RT_b^{(0)}
\frac{R_a^{add}}{\epsilon}+R_aR_b +{\cal O}(\epsilon).
\end{equation}
   Expanding $R_aR_b$ in $\epsilon$ and simplifying the resulting expression gives
\begin{equation}
    \IT_a \IT_b -\frac{I_a^{add}I_b^{add}}{\epsilon^2} +\frac{R_a^{add}R_b^{add}}{\epsilon^2}
+\frac{H_a^{UV}H_b^{UV}}{\epsilon^2}
+\frac{H_a^{UV}\RT_b^{(0)}+\RT_a^{(0)}H_b^{UV}}{\epsilon}
+(R_aR_b)^{(0)} + {\cal O}(\epsilon),
\end{equation}
where $(R_a R_b)^{(0)}$ stands for the terms proportional to
$\epsilon^0$ in the product $R_aR_b$.
   Using again $I_i^{add}=-R_i^{add}$ we see that all terms
containing the additional divergences vanish,
\begin{eqnarray}
    && \IT_a\IT_b +\frac{H_a^{UV}H_b^{UV}}{\epsilon^2}
+\frac{H_a^{UV}\RT_b^{(0)}+\RT_a^{(0)}H_b^{UV}}{\epsilon}
+(R_aR_b)^{(0)} + {\cal O}(\epsilon)\,.
\end{eqnarray}
   The term $R_aR_b$ satisfies the Ward identities, in particular
each term in the $\epsilon$ expansion of $R_aR_b$ does so
individually.
   This means that we can subtract the finite part of $R_aR_b$ by
a counterterm.
   The terms proportional to $\epsilon^{-2}$ and $\epsilon^{-1}$
stem from the UV divergences in $H_a$ and $H_b$.
   These terms also satisfy the
Ward identities individually and are absorbed in counterterms.
   As desired, the renormalized result for the product of two one-loop
integrals including the counterterm integrals is then simply the
product of the renormalized one-loop integrals,
\begin{equation}\label{Ren:ProductResult}
    (H_a H_b)^{r}=\IT_a\IT_b.
\end{equation}

   Besides respecting all symmetries the renormalization
prescription must also result in a proper power counting for
renormalized integrals.
   The  chiral order of a product of two integrals is the
sum of the individual orders.
   For a one-loop integral the infrared
singular part $\IT$ satisfies the power counting.
   Therefore the result of Eq.~(\ref{Ren:ProductResult}) also
satisfies power counting.

\section{Infrared renormalization of two-loop integrals relevant
to the nucleon mass calculation}\label{Sec:TwoLoop}

   In this section we describe the renormalization procedure
for two-loop integrals that do not directly factorize into the
product of two one-loop integrals.
   We follow the general method presented in
Ref.~\cite{Schindler:2003je}, but give more details.
   First we show how the proper renormalization of two-loop
integrals and of the corresponding counterterm integrals preserves
the underlying symmetries.
   Next we describe a simplified formalism to
arrive at the same results while greatly reducing the
calculational difficulties.

\subsection{General method}\label{Ren:Concept}

   Denote a general two-loop integral contributing to the nucleon mass by $H_2$,
\begin{equation}\label{Ren:2loopIntDef}
    H_2(a,b,c,d,e|n)=\iint \frac{d^nk_1 d^nk_2}{(2\pi)^{2n}} \, \frac{1}{A^a B^b C^c D^d
    E^e}\,,
\end{equation}
where
\begin{eqnarray}\label{Ren:Denom}
  A &=& k_1^2-M^2+i0^+, \nonumber\\
  B &=& k_2^2-M^2+i0^+, \nonumber\\
  C &=& k_1^2+2p\cdot k_1+i0^+, \nonumber\\
  D &=& k_2^2+2p\cdot k_2+i0^+, \nonumber\\
  E &=& k_1^2+2p\cdot k_1+2k_1\cdot k_2+2p\cdot k_2+k_2^2+i0^+.
\end{eqnarray}
   Using a dimensional counting analysis we can write $H_2(a,b,c,d,e|n)$ as\footnote{For brevity we
employ the notation $H_2$ for  $H_2(a,b,c,d,e|n)$ in the following
discussion.}
\begin{equation}\label{Ren:DimCount}
    H_2=F_1+F_2+F_3+F_4.
\end{equation}
   $F_1$ is obtained by simply expanding the integrand in $M$
and interchanging summation and integration.
   For $F_2$ we rescale the first loop momentum $k_1$ by
\begin{equation}\label{Ren:k1Rescale}
    k_1 \mapsto \frac{M}{m}\, k_1,
\end{equation}
expand the resulting integrand in $M$, and interchange summation
and integration.
   $F_3$ is obtained analogously to $F_2$, only that instead of
$k_1$ the second loop momentum $k_2$ is rescaled,
\begin{equation}\label{Ren:k2Rescale}
    k_2 \mapsto \frac{M}{m}\, k_2.
\end{equation}
   Finally $F_4$ is defined as the result from simultaneously
rescaling both loop momenta,
\begin{equation}\label{Ren:k1k2Rescale}
    k_1 \mapsto \frac{M}{m}\, k_1, \quad k_2 \mapsto \frac{M}{m}\, k_2,
\end{equation}
and expanding the integrand with subsequent interchange of
summation and integration.
   $F_1$, $F_2+F_3$, and $F_4$ separately satisfy the Ward
identities due to different overall factors of $M$.
   This is analogous to the one-loop sector, where the infrared
singular and infrared regular parts separately satisfy the Ward
identities, since the infrared singular part is nonanalytic in
small quantities for noninteger $n$, while the infrared regular
term is analytic.
   As in the one-loop case the interchange of summation and
integration generates additional divergences not present in $H_2$ in
each of the terms $F_1$, $F_2+F_3$, and $F_4$.
   Again, these additional divergences cancel in the sum of all terms.

   In addition to the two-loop integral we also need to determine
the corresponding subintegrals.
   To identify the first subintegral we consider the $k_1$
integration in $H_2$,
\begin{equation}\label{Ren:Sub1}
    H_{sub_1} = \int \frac{d^nk_1}{(2\pi)^{n}} \, \frac{1}{A^a C^c E^e}\,.
\end{equation}
   This is a one-loop integral which is renormalized using
``standard'' infrared renormalization.
   The infrared regular part $R_{sub_1}$ of this integral is obtained
by expanding the integrand in $M$ and interchanging summation and
integration.
   The only term in Eq.~(\ref{Ren:Sub1}) depending on $M$ is $A$.
   Symbolically we write
\begin{equation}\label{Ren:RSub1}
    R_{sub_1} = \sum \int \frac{d^nk_1}{(2\pi)^{n}} \, \frac{1}{\underline{A}^a C^c
    E^e}\,,
\end{equation}
where underlined expressions are understood as an expansion in
$M$.
   $R_{sub_1}$ contains additional divergences, and we define
$\widetilde{R}_{sub_1}$ as $R_{sub_1}$ without these
divergences,\footnote{Note that for the integrals of interest
here, the UV divergence is included in the infrared regular part
$R$.}
\begin{equation}\label{Ren:RSub1tilde}
   \widetilde{R}_{sub_1} =
   R_{sub_1}-\frac{R_{sub_1}^{add}}{\epsilon}\,.
\end{equation}
   As in the definition of Eq.~(\ref{IR:Hepsilon}), $\widetilde{R}_{sub_1}$ again
contains all terms of positive powers of $\epsilon$.
   Since $H_{sub_1}$ is a standard one-loop integral, $\widetilde{R}_{sub_1}$ will satisfy the Ward
identities and can be absorbed in counterterms of the Lagrangian.

   Using these counterterms as a vertex we obtain a counterterm
integral of the form
\begin{equation}\label{Ren:CT1}
    H_{CT_1} = - \int\frac{d^nk_2}{(2\pi)^n}\, \widetilde{R}_{sub_1}
    \frac{1}{B^b D^d}\,.
\end{equation}
   $H_{CT_1}$ is generated by a Lagrangian that is consistent
with the considered symmetries.
   Therefore, $H_{CT_1}$ satisfies the Ward identities.
   Inserting Eqs.~(\ref{Ren:RSub1}) and (\ref{Ren:RSub1tilde}) we rewrite $H_{CT_1}$ as
\begin{equation}\label{Ren:CT1Detail}
    H_{CT_1} = - \int\frac{d^nk_2}{(2\pi)^n}\,\sum\int\frac{d^nk_1}{(2\pi)^n} \frac{1}{\underline{A}^a B^b C^c D^d E^e}
+\int\frac{d^nk_2}{(2\pi)^n}\,
\frac{R_{sub_1}^{add}}{\epsilon}\frac{1}{B^b D^d}\,.
\end{equation}
   Equation~(\ref{Ren:CT1Detail}) still needs to be renormalized.
   After the $k_1$ integration has been performed,
Eq.~(\ref{Ren:CT1Detail}) is a one-loop integral and standard
infrared renormalization can be used.
   To obtain the infrared singular part $I_{CT_1}$ we rescale
$k_2 \mapsto \frac{M}{m}\, k_2$, expand in $M$, and interchange
summation and integration.
   Symbolically we write
\begin{equation}\label{Ren:ICT1}
    I_{CT_1} = - \sum \int\frac{d^nk_2}{(2\pi)^n}\,
\sum\int\frac{d^nk_1}{(2\pi)^n} \frac{1}{\underline{A}^a
\underline{\underline{B}}^b C^c \underline{\underline{D}}^d
\underline{\underline{E}}^e} +\sum\int\frac{d^nk_2}{(2\pi)^n}\,
\epsilon^{-1}\underline{\underline{R_{sub_1}^{add}}}\;\frac{1}{\underline{\underline{B}}^b
\underline{\underline{D}}^d}\,,
\end{equation}
   where double-underlined quantities are first rescaled and then
expanded.
   Note that $R_{sub_1}^{add}$ can also depend on $k_2$ through the
denominator $E$ in Eq.~(\ref{Ren:Sub1}).
   Since $I_{CT_1}$ is obtained from a one-loop integral that satisfies
the Ward identities through the standard infrared renormalization
process, it will itself satisfy the Ward identities.
   The infrared renormalized expression for the counterterm
integral,
\begin{equation}\label{Ren:ICT1Tilde}
   \widetilde{I}_{CT_1} = I_{CT_1} -
   \frac{I_{CT_1}^{add}}{\epsilon}\,,
\end{equation}
then also satisfies the Ward identities.
   Note that $I_{CT_1}^{add}$ itself contains terms proportional to
$\epsilon^{-1}$, since it stems from the one-loop counterterm for
the subintegral, but we choose not to include any terms
proportional to positive powers of $\epsilon$. This means that
$\epsilon^{-1}I_{CT_1}^{add}$ only contains terms proportional to
$\epsilon^{-2}$ and $\epsilon^{-1}$. The expression for
$\widetilde{I}_{CT_1}$ therefore does not contain any divergent
terms stemming from additional divergences.

   We now show how $\widetilde{I}_{CT_1}$ is related to the term
$F_3$ of Eq.~(\ref{Ren:DimCount}).
   As explained above, $F_3$ is obtained by rescaling $k_2$,
expanding the resulting integrand and interchanging summation and
integration.
   In the above notation this would correspond to
\begin{equation}\label{Ren:F3Notation}
   F_3 = \sum\iint \frac{d^nk_1d^nk_2}{(2\pi)^{2n}}\,
\frac{1}{\underline{A}^a \underline{\underline{B}}^b C^c
\underline{\underline{D}}^d \underline{\underline{E}}^e}\,.
\end{equation}
   Comparing with the first term in Eq.~(\ref{Ren:ICT1}) we see that the integrands in
both cases are expanded in the same way.
   Therefore, when adding the counterterm diagram $\widetilde{I}_{CT_1}$ to
$H_2$ it cancels parts of $F_3$.
   The difference between $\widetilde{I}_{CT_1}$ and $F_3$ is that in $F_3$
the terms stemming from the additional divergences
$R_{sub_1}^{add}$ (including \emph{finite} terms) as well as the
additional divergences $I_{CT_1}^{add}/\epsilon$ that are
proportional to $\epsilon^{-2}$ and $\epsilon^{-1}$ are not
subtracted.
   As pointed out above, the original integral $H_2$ only contains
UV divergences, therefore the additional divergences cancel in the
sum $F_1+F_2+F_3+F_4$.
   Apart from these contributions, the terms remaining in the sum
$\widetilde{I}_{CT_1}+F_3$ are the finite contributions stemming
from the additional divergences in $R_{sub_1}$.
   Since in $F_3$ the variable $k_2$ is rescaled before expanding
while the $k_1$ variable remains unchanged, $F_3$ can be
considered as a sum of products of infrared singular and infrared
regular terms, which we symbolically write as
\begin{equation}\label{Ren:F3Product}
    F_3=\sum R_1 I_2\,.
\end{equation}
   In this notation the remaining finite terms are $\sum R_1^{add}I_2^{(1)}$,
where $\epsilon^{-1}R_1^{add}$ is the additional divergence of
$R_1$ and $I_2^{(1)}$ is the part of $I_2$ proportional to
$\epsilon$.

   The second subdiagram can be calculated analogously, and
is related to the term $F_2$ in Eq.~(\ref{Ren:DimCount}).

   Taking the above considerations into account we obtain for the sum
of the original integral $H_2$ and the corresponding counterterm
integrals
\begin{eqnarray}\label{Ren:H+CT}
  H_2+\widetilde{I}_{CT_1}+\widetilde{I}_{CT_2} &=&
F_1+F_2+F_3+F_4+\widetilde{I}_{CT_1}+\widetilde{I}_{CT_2} \nonumber \\
  &=& \widetilde{F}_1+\widetilde{F}_4+ \sum
R_1^{add}I_2^{(1)}+\sum R_2^{add}I_1^{(1)}\nonumber\\
  &=& \widetilde{F}_1+\widetilde{F}_4 - \sum I_1^{add}I_2^{(1)}
-\sum I_2^{add}I_1^{(1)}\,,
\end{eqnarray}
where $\widetilde{F}_i$ indicates that the additional divergences
are excluded.

   The expression in Eq.~(\ref{Ren:H+CT}) satisfies the Ward
identities since each term in the sum on the left side of the
first line does so individually.
   $F_1$ separately satisfies the Ward identities, in
particular this is the case for each term in its $\epsilon$
expansion.
   This means that we can subtract the finite part of
$\widetilde{F}_1$ by an overall counterterm without violating the
symmetries.
   Since the remaining UV divergences also satisfy the Ward
identities, absorbing them in an overall counterterm does not
violate the symmetries.
   The result for the renormalized two-loop diagram is then
\begin{equation}\label{Ren:I}
    H_2^r=\widetilde{F}_4- \sum I_1^{add}I_2^{(1)}
-\sum I_2^{add}I_1^{(1)}\,.
\end{equation}
   Since all subtractions preserve the symmetries $H_2^r$ will satisfy
the Ward identities.

   So far we have subtracted pole parts in the epsilon expansion.
   Following \cite{Becher:1999he} we choose to absorb the
combination
\begin{equation}\label{Ren:lambda}
    \frac{1}{(4\pi)^2}\,\left[ \frac{1}{n-4}\,-\frac{1}{2}\,
    \left(\log(4\pi)-\gamma_E+1\right) \right]
\end{equation}
instead, which is achieved by simply replacing the original t'Hooft
parameter $\tilde\mu$ by
\begin{equation}\label{Ren:MSTilde}
    \tilde\mu \to \frac{\mu}{(4\pi)^{1/2}}\,
    e^{\frac{\gamma_E-1}{2}}\,
\end{equation}
(see also App.~\ref{App:IntEv}).

   $F_4$ is obtained by rescaling both $k_1$ and $k_2$ and satisfies
the power counting.
   Since the terms $I_i$ result from the rescaling of $k_i$, the
product $I_1 I_2$ has the same analytic structure in $M$ as $F_4$,
and therefore satisfies the power counting.
   This means that also the renormalized integral
$H_2^r=\widetilde{F}_4- \sum I_1^{add}I_2^{(1)} -\sum
I_2^{add}I_1^{(1)}$ obeys the power counting.

\subsection{Simplified method}

   In the previous subsection we have established the concept of
infrared renormalization of two-loop integrals.
   The procedure outlined above is quite involved when
applied to actual calculations of physical processes.
   Therefore, we now describe a simpler method of obtaining
the renormalized expression $H_2^r$ which, however, is only
applicable to integrals with a \emph{single} small scale.
   This is the case for the calculation of the nucleon mass,
whereas e.g. the nucleon form factors contain the momentum
transfer as an additional small quantity.

   Instead of calculating the subintegrals of the original
integral $H_2$, consider just the terms in $F_4$.
   $F_4$ itself is a sum of two-loop integrals.
   Each two-loop integral contains one-loop subintegrals, which
correspond to performing only one loop integration while keeping
the other one fixed.
   These subintegrals contain divergences, resulting in
divergent as well as finite contributions when the second loop
integration is performed.
   In addition to the subintegral contributions, $F_4$ contains finite parts
and additional divergences originating in the interchange of
summation and integration when generating $F_4$.
   We can symbolically write $F_4$ as
\begin{equation}\label{Ren:F4Parts}
    F_4 =
    \bar{F}_4+\frac{\bar{F}_4^{add,2}}{\epsilon^2}+\frac{\bar{F}_4^{add,1}}{\epsilon}
    +\frac{F_4^{Sub_1,div}}{\epsilon}\,F_4^{k_2}+\frac{F_4^{Sub_2,div}}{\epsilon}\,F_4^{k_1}\,.
\end{equation}
   Here, the finite parts of $F_4$ are denoted by $\bar{F}_4$ to
distinguish them from $\widetilde{F}_4$ in Eq.~(\ref{Ren:H+CT}).
   The bar notation is also used for the divergent terms $\bar{F}_4^{add,2}$
and $\bar{F}_4^{add,1}$ to show that these are not the complete
divergent expressions for $F_4$, but only the additional
divergences of order $\epsilon^{-2}$ and $\epsilon^{-1}$,
respectively.
   The terms $\epsilon^{-1}F_4^{Sub_i,div}$ denote the
divergences of the subintegral with respect to the integration
over $k_i$, while $F_4^{k_j}$ stands for the remaining second
integration of the counterterm integral.
   Note that the divergent part of the first loop integration over
$k_i$ in general depends on the second loop momentum $k_j$.
   This dependence is included in the expression $F_4^{k_j}$.

   We now show how the different parts in Eq.~(\ref{Ren:F4Parts})
are related to expressions in $F_2$ and $F_3$ and then describe
the simplified renormalization method.
   $F_4$ is obtained from the original integral $H$ by rescaling
$k_1$ and $k_2$, expanding the resulting integrand in $M$ and
interchanging summation and integration.
   For the denominators of Eq.~(\ref{Ren:Denom}) the rescaling results
in
\begin{eqnarray*}
  k_1^2-M^2+i0^+ &\mapsto& \left(\frac{M}{m}\right)^2(k_1^2-m^2+i0^+), \\
  k_2^2-M^2+i0^+ &\mapsto& \left(\frac{M}{m}\right)^2(k_2^2-m^2+i0^+), \\
  k_1^2+2p\cdot k_1+i0^+ &\mapsto& \left(\frac{M}{m}\right)\left(\frac{M}{m}\, k_1^2+2p\cdot k_1+i0^+\right), \\
  k_2^2+2p\cdot k_2+i0^+ &\mapsto& \left(\frac{M}{m}\right)\left(\frac{M}{m}\, k_2^2+2p\cdot k_2+i0^+\right), \\
  k_1^2+2p\cdot k_1+2k_1\cdot k_2+2p\cdot k_2+k_2^2+i0^+ &\mapsto& \\
  && \hspace{-12em}\left(\frac{M}{m}\right)\left(\frac{M}{m}\, k_1^2+2p\cdot k_1+2\frac{M}{m}\, k_1\cdot k_2
  +2p\cdot k_2+\frac{M}{m}\,k_2^2+i0^+\right).
\end{eqnarray*}
After the interchange of summation and integration one can perform
the substitution $k_i \mapsto \frac{m}{M}\,k_i$ to bring the
denominators $k_1^2-m^2+i0^+$ and $k_2^2-m^2+i0^+$ back into the
form $A$ and $B$, respectively.
   The result can be interpreted as obtained from the original
integral by leaving $A$ and $B$ unchanged and expanding $C$ in
$k_1^2$, $D$ in $k_2^2$, and $E$ in $k_1^2+2k_1\cdot k_2+k_2^2$,
respectively.
   Symbolically
\begin{eqnarray}\label{Ren:F4exp}
    F_4 \hspace{-.5em}&\sim& \hspace{-.5em}
\sum \iint
\frac{d^nk_1d^nk_2}{(2\pi)^{2n}}\,\frac{1}{[k_1^2-M^2+i0^+]^a
[k_2^2-M^2+i0^+]^b [\underline{k_1}^2+2p\cdot k_1+i0^+]^c}  \nonumber\\
   &&\times \frac{1}{[\underline{k_2}^2+2p\cdot k_2+i0^+]^d
[\underline{k_1}^2+2p\cdot k_1+\underline{2k_1\cdot k_2}+2p\cdot
k_2+\underline{k_2}^2+i0^+]^e}\,,\nonumber\\
\end{eqnarray}
where we have used the underlined notation to mark terms that we
have expanded in.

   The divergent parts of the $k_1$ subintegral stem from
the integration region $k_1\to \infty$.
   They can be generated by further expanding each term in $F_4$
in \emph{inverse} powers of $k_1$.
   This corresponds to an expansion in \emph{positive} powers of
$M$ for the first denominator and in \emph{positive} powers of
$2p\cdot k_2$ in the resulting last propagator,
   \begin{align}\label{Ren:F4sub1div}
& \frac{F_4^{Sub_1,div}}{\epsilon}\,F_4^{k_2} \sim \sum \iint
\frac{d^nk_2d^nk_1}{(2\pi)^{2n}}\,\frac{1}{[k_1^2-\underline{M}^2+i0^+]^a
[k_2^2-M^2+i0^+]^b}   \notag\\
   &\quad \times \frac{1}{[\underline{k_1}^2+2p\cdot k_1+i0^+]^c[\underline{k_2}^2+2p\cdot
   k_2+i0^+]^d} \notag \\
   &\quad \times \frac{1}{[\underline{k_1}^2+2p\cdot k_1+\underline{2k_1\cdot
k_2}+\underline{2p\cdot k_2}+\underline{k_2}^2+i0^+]^e}\,.
\end{align}
   We see that the expression for $F_4^{k_2}$ is of
the form
\begin{equation}\label{Ren:HF4k2}
    F_4^{k_2} \sim \sum\int\frac{d^nk_2}{(2\pi)^n}\,
\frac{f_{\mu\nu\lambda\cdots}\,k_2^\mu k_2^\nu
k_2^\lambda\cdots}{[k_2^2-M^2+i0^+]^b[2p\cdot
k_2+i0^+]^{d+i_1}}\,,
\end{equation}
where $f_{\mu\nu\lambda\cdots}$ denotes the coefficients that
result from the expansion in Eq.~(\ref{Ren:F4sub1div}).

   Next we show that $F_4^{k_2}$ is related to terms in
$F_3$.
   $F_3$ is generated from the original integral $H_2$ by rescaling
$k_2$, expanding the resulting integrand, and interchanging
summation and integration.
   After the substitution $k_2 \mapsto \frac{m}{M}\,k_2$ and using the above notation we write
\begin{eqnarray}\label{Ren:F3exp}
    F_3 \hspace{-.5em}&\sim& \hspace{-.5em}
\sum \iint \frac{d^nk_1
d^nk_2}{(2\pi)^{2n}}\,\frac{1}{[k_1^2-\underline{M}^2+i0^+]^a
[k_2^2-M^2+i0^+]^b [k_1^2+2p\cdot k_1+i0^+]^c}  \nonumber\\
   && \times \frac{1}{[\underline{k_2}^2+2p\cdot k_2+i0^+]^d
[k_1^2+2p\cdot k_1+\underline{2k_1\cdot k_2}+\underline{2p\cdot
k_2}+\underline{k_2}^2+i0^+]^e}
\nonumber\\
    \hspace{-.5em}&\sim& \hspace{-.5em}
\sum \iint \frac{d^nk_1 d^nk_2}{(2\pi)^{2n}}\,\frac{[k_2^2]^{j_4}
[2k_1\cdot k_2]^{j_5}} {[k_1^2+i0^+]^{a+j_1}
[k_1^2+2p\cdot k_1+i0^+]^{c+e+j_2}[k_2^2-M^2+i0^+]^b}\nonumber\\
   && \times \frac{1}{[2p\cdot k_2+i0^+]^{d+j_3}}\,.
\end{eqnarray}
   We see that $F_3$ is the sum of products of one-loop
(tensorial) integrals.
   As explained above these products of one-loop
integrals are in fact products of infrared singular and infrared
regular parts of integrals (see Eq.~(\ref{Ren:F3Product})),
\begin{displaymath}
    F_3=\sum R_1 I_2\,,
\end{displaymath}
    and the expressions for $I_2$ are given by
\begin{equation}\label{Ren:F3I2}
    I_2 \sim \sum\int\frac{d^nk_2}{(2\pi)^n}\,
\frac{k_2^\alpha k_2^\beta
k_2^\gamma\cdots}{[k_2^2-M^2+i0^+]^b[2p\cdot k_2+i0^+]^{d+i_2}}\,.
\end{equation}
   Considering the $k_2$ integrals of Eqs.~(\ref{Ren:F4sub1div}) and (\ref{Ren:F3exp})
one sees that one has expanded in the same quantities.
   While the ordering of the expansions as well as the interchanges
of summation and integration are different, the two expansions are
equivalent.
    Therefore, comparing Eqs.~(\ref{Ren:HF4k2}) and (\ref{Ren:F3I2}), one finds that
for each term in $F_4^{k_2}$ there is a corresponding term in
$I_2$, or symbolically
\begin{equation}\label{Ren:EqualityHF4k2}
   F_4^{k_2} = I_2\,.
\end{equation}
   An analogous analysis for the second subintegral gives
\begin{equation}\label{Ren:EqualityHF4k1}
   F_4^{k_1} = I_1\,.
\end{equation}

   As a next step we show that the divergences of the $F_4$
subintegrals are related to the additional divergences of the
integrals $R_i$ in $F_2$ and $F_3$.
   From Eq.~(\ref{Ren:F4sub1div}) we see that the divergent part of the
$k_1$ subintegral is given by integrals of the type
\begin{eqnarray}\label{Ren:F4sub1Int}
    F_4^{Sub_1} &\sim& \sum \int
    \frac{d^nk_1}{(2\pi)^{n}}\,\frac{1}{[k_1^2-M^2+i0^+]^a
[\underline{k_1}^2+2p\cdot k_1+i0^+]^c}\nonumber\\
   && \times \quad\frac{1}{
[\underline{k_1}^2+2p\cdot k_1+\underline{2k_1\cdot
k_2}+\underline{2p\cdot k_2}
+\underline{k_2}^2+i0^+]^e}\nonumber\\
   &\sim& \sum \int\frac{d^nk_1}{(2\pi)^{n}}\, \frac{k_1^\mu k_1^\nu\cdots}
{[k_1^2-M^2+i0^+]^{a}[2p\cdot k_1+i0^+]^{c+e+l_2}}\,.
\end{eqnarray}
   The infrared regular integrals $R_1$ in Eq.~(\ref{Ren:F3exp}) read
\begin{equation}\label{Ren:R1}
    R_1 \sim \sum\int \frac{d^nk_1}{(2\pi)^{n}}\, \frac{k_1^\mu k_1^\nu\cdots}
{[k_1^2+i0^+]^{a+m_1}[k_1^2+2p\cdot k_1+i0^+]^{c+e+m_2}}\,.
\end{equation}
   $F_4^{Sub_1}$ and $R_1$ can be interpreted as the infrared
singular and infrared regular parts of the auxiliary integrals
\begin{equation}\label{Ren:MotherInt}
    h \sim \!\sum\!\int\! \frac{d^nk_1}{(2\pi)^{n}}\, \frac{k_1^\mu k_1^\nu\cdots}
{[k_1^2-M^2+i0^+]^{\alpha}[k_1^2+2p\!\cdot\! k_1+i0^+]^{\beta}}\,,
\end{equation}
respectively.
   Since $h$ is a ``standard'' one-loop integral that is only UV divergent,
the additional divergences in its IR regular part $R_1$ must
cancel exactly with the divergences in its IR singular part
$F_4^{Sub_1,div}$.
   Therefore,
\begin{equation}\label{Ren:DivEq1a}
    \frac{F_4^{Sub_1,div}}{\epsilon} = -\frac{R_1^{add}}{\epsilon}\,,
\end{equation}
and, using $R_1^{add}=-I_1^{add}$, it also follows that
\begin{equation}\label{Ren:DivEq1b}
    \frac{F_4^{Sub_1,div}}{\epsilon} =
    \frac{I_1^{add}}{\epsilon}\,.
\end{equation}
   Analogously
\begin{equation}\label{Ren:DivEq2}
    \frac{F_4^{Sub_2,div}}{\epsilon} = -\frac{R_2^{add}}{\epsilon}
    = \frac{I_2^{add}}{\epsilon}\,.
\end{equation}

   Having established the relationship between the terms in
$F_4$ and the terms in $F_2$ and $F_3$ we now describe the
renormalization procedure.
   Our method consists of treating each two-loop integral
contributing to $F_4$ as an independent integral.
   We then renormalize each two-loop integral in the $\widetilde{\mbox{MS}}$
scheme, i.e. we
\begin{itemize}
    \item[-] determine the divergences in the subintegrals,
    \item[-] use the divergences as vertices in one-loop counterterm
    integrals that are added to $F_4$,
    \item[-] perform an additional overall subtraction by absorbing
    all remaining divergences in counterterms,
    \item[-] replace $\tilde\mu = \frac{\mu}{(4\pi)^{1/2}}\,
    e^{\frac{\gamma_E-1}{2}}$ and set $\mu=m$.
\end{itemize}
   The divergences in the subintegrals are given by
$\epsilon^{-1}F_4^{Sub_i,div}$.
   The one-loop counterterm integrals using these divergences read
\begin{equation}\label{Ren:RecipeCT}
    -\frac{F_4^{Sub_1,div}}{\epsilon}\,F_4^{k_2}-\frac{F_4^{Sub_2,div}}{\epsilon}\,F_4^{k_1}\,.
\end{equation}
   According to Eqs.~(\ref{Ren:EqualityHF4k2}), (\ref{Ren:EqualityHF4k1}),
(\ref{Ren:DivEq1b}), and (\ref{Ren:DivEq2}) this can be written as
\begin{equation}\label{Ren:RecipeCTReplaced}
    -\frac{I_2^{add}}{\epsilon}\,I_1-\frac{I_1^{add}}{\epsilon}\,I_2\,.
\end{equation}
   When added to $F_4$ we obtain
\begin{equation}\label{Ren:RecipeF4Renorm}
    F_4-\frac{I_2^{add}}{\epsilon}\,I_1-\frac{I_1^{add}}{\epsilon}\,I_2\,.
\end{equation}
   Using the notation of Subsec.~\ref{Ren:Concept}, we write $F_4$ as
the sum of the additional divergences and a remainder
$\widetilde{F}_4$,
\begin{equation}\label{Ren:RecipeF4Parts}
    F_4=\frac{F_4^{add,2}}{\epsilon^2}+\frac{F_4^{add,1}}{\epsilon}+\widetilde{F}_4\,.
\end{equation}
   Note that the divergent terms $F_4^{add,i}$ are not the
divergent expressions $\bar{F}_4^{add,i}$ of
Eq.~(\ref{Ren:F4Parts}).
   Performing the $\epsilon$ expansion for the integrals $I_i$,
$$ I_i=\epsilon^{-1}I_i^{add}+ I_i^{(0)}+\epsilon I_i^{(1)}+\cdots\,,$$
the sum of $F_4$ and the counterterm integrals is given by
\begin{equation}\label{Ren:RecipeF4RenormExp}
   \frac{F_4^{add,2}}{\epsilon^2}+\frac{F_4^{add,1}}{\epsilon}
-2\frac{I_2^{add}I_1^{add}}{\epsilon^2}-\frac{I_1^{add}}{\epsilon}\,I_2^{(0)}
-\frac{I_2^{add}}{\epsilon}\,I_1^{(0)} +\widetilde{F}_4-I_1^{add}
I_2^{(1)}-I_2^{add} I_1^{(1)}\,.
\end{equation}

   We now show that the remaining divergences are analytic in
$M^2$ and can therefore be absorbed by counterterms.
   Recall that the sum of all additional divergences has to vanish,
since they are not present in the original integral,
\begin{equation}\label{Ren:SumIRDiv}
    0 = \frac{F_1^{add,2}}{\epsilon^2}+\frac{F_1^{add,1}}{\epsilon}
+\frac{F_2^{add,2}}{\epsilon^2}+\frac{F_2^{add,1}}{\epsilon}
+\frac{F_3^{add,2}}{\epsilon^2}+\frac{F_3^{add,1}}{\epsilon}
+\frac{F_4^{add,2}}{\epsilon^2}+\frac{F_4^{add,1}}{\epsilon}\,.
\end{equation}
   As shown above $F_2$ and $F_3$ are the sums of products of
one-loop integrals, so Eq.~(\ref{Ren:SumIRDiv}) can be rewritten
as
\begin{eqnarray}
  0 &=& \frac{F_1^{add,2}}{\epsilon^2}+\frac{F_1^{add,1}}{\epsilon}
+\frac{I_1^{add}R_2^{add}}{\epsilon^2}+\frac{I_1^{add}}{\epsilon}\,R_2^{(0)}
+I_1^{(0)}\frac{R_2^{add}}{\epsilon}
+\frac{I_2^{add}R_1^{add}}{\epsilon^2}+\frac{I_2^{add}}{\epsilon}\,R_1^{(0)}\nonumber\\
  &&
+I_2^{(0)}\frac{R_1^{add}}{\epsilon}
+\frac{F_4^{add,2}}{\epsilon^2}+\frac{F_4^{add,1}}{\epsilon}\,.
\end{eqnarray}
  Making use of $I_i^{add}=-R_i^{add}$ the sum of all additional
divergences takes the form
\begin{eqnarray}\label{Ren:SumIRDivAnalytic}
  0 &=& \frac{F_1^{add,2}}{\epsilon^2}+\frac{F_1^{add,1}}{\epsilon}
-\frac{R_1^{add}}{\epsilon}\,R_2^{(0)}
-\frac{R_2^{add}}{\epsilon}\,R_1^{(0)}\nonumber\\
  &&-2\frac{I_1^{add}I_2^{add}}{\epsilon^2}
-\frac{I_1^{add}}{\epsilon}\,I_2^{(0)}-\frac{I_2^{add}}{\epsilon}\,I_1^{(0)}
+\frac{F_4^{add,2}}{\epsilon^2}+\frac{F_4^{add,1}}{\epsilon}\,.
\end{eqnarray}
   All terms in $F_1$ for the two-loop integral as well as the
infrared regular terms in one-loop integrals are analytic in
$M^2$.
   Therefore the first line in Eq.~(\ref{Ren:SumIRDivAnalytic}) is
analytic in $M^2$.
   Since the sum of all terms vanishes the second line also has
to be analytic.
   This second line, however, comprises exactly the remaining divergences in
Eq.~(\ref{Ren:RecipeF4RenormExp}), which are therefore analytic in
$M^2$ and can be subtracted.
   After these divergences have been absorbed in counterterms, the resulting expression for the
renormalized contribution of $F_4$ reads
\begin{equation}\label{Ren:RecipeF4RenormFinal}
   F_4^r=\widetilde{F}_4-\sum I_1^{add}
I_2^{(1)}-\sum I_2^{add} I_1^{(1)}\,,
\end{equation}
where we have explicitly shown the sums again.
   Comparing with Eq.~(\ref{Ren:I}) we see that our result exactly
reproduces the expression for the renormalized original integral
$H_2^r$.

\subsection{$\epsilon$-dependent factors}

   For actual calculations it is often convenient to reduce
tensorial integrals to scalar integrals before performing the
dimensional counting analysis as well as the renormalization.
   The reduction of the tensorial integrals can result in $\epsilon$-dependent
factors multiplying the scalar integrals.
   These change the form of the result of Eq.~(\ref{Ren:I}) since additional
finite terms can appear.
   Let the $\epsilon$-dependent factor be given by
\begin{equation}\label{Ren:phi}
    \phi(\epsilon)=\phi^{(0)} + \epsilon \phi^{(1)} + \epsilon^2 \phi^{(2)} +
    \cdots.
\end{equation}
   Consider performing the $k_1$ integration first.
   Suppose that from the result one can extract an $\epsilon$-dependent factor
$\varphi_1(\epsilon)$, and the subsequently performed $k_2$
integration leads to another $\epsilon$-dependent factor,
$\varphi_2(\epsilon)$, with
\begin{equation}\label{Ren:phiProduct1}
    \phi(\epsilon)=\varphi_{1}(\epsilon) \cdot \varphi_{2}(\epsilon).
\end{equation}
   One can also perform the $k_2$ integration first, which leads
to a different factor $\tilde\varphi_2(\epsilon)$, followed by the
$k_1$ integration resulting in a factor
$\tilde\varphi_1(\epsilon)$ with
\begin{equation}\label{Ren:phiProduct2}
    \phi(\epsilon)=\tilde\varphi_{2}(\epsilon) \cdot
    \tilde\varphi_{1}(\epsilon)\,.
\end{equation}
   The terms $\varphi_1(\epsilon)=\varphi_1^{(0)} + \epsilon
   \varphi_1^{(1)}+\epsilon^2\varphi_1^{(2)} +\cdots$ and
$\tilde\varphi_2(\epsilon)=\tilde\varphi_2^{(0)} + \epsilon
\tilde\varphi_2^{(1)} +\epsilon^2\tilde\varphi_2^{(2)} +\cdots$
can then directly be taken into account when determining the
divergent contributions from subintegrals.
   The result $H_2^{r,\,\phi}$ for the renormalized integral $\phi(\epsilon)H_2$
reads
\begin{equation}\label{Ren:phiI}
    H_2^{r,\,\phi} = \widetilde{F}_4^\phi
-\varphi_1^{(0)} I_1^{add}\left(\varphi_2^{(2)} I_2^{add}+
\varphi_2^{(1)} I_2^{(0)}+\varphi_2^{(0)} I_2^{(1)}\right)
-\tilde\varphi_2^{(0)} I_2^{add}\left(\tilde\varphi_1^{(2)}
I_1^{add}+\tilde\varphi_1^{(1)} I_1^{(0)}+\tilde\varphi_1^{(0)}
I_1^{(1)}\right),
\end{equation}
where $\widetilde{F}_4^\phi$ denotes the finite terms in
$\phi(\epsilon)\, F_4$, and $I_i^{(0)}$, $\varphi_1^{(0)}$ and
$\tilde\varphi_2^{(0)}$ are the $\epsilon$-independent terms in
$I_i$, $\varphi_1$ and $\tilde\varphi_2$, respectively.
    Our simplified method still holds provided the $\epsilon$-dependent
factors are taken into account.

\begin{figure}
\begin{center}
\epsfig{file=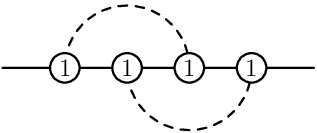,width=0.3\textwidth}
\end{center}
\caption{Two-loop diagram contributing to the nucleon
self-energy.\label{Ren:TwoLoopDia}}
\end{figure}

\begin{figure}
\begin{center}
\epsfig{file=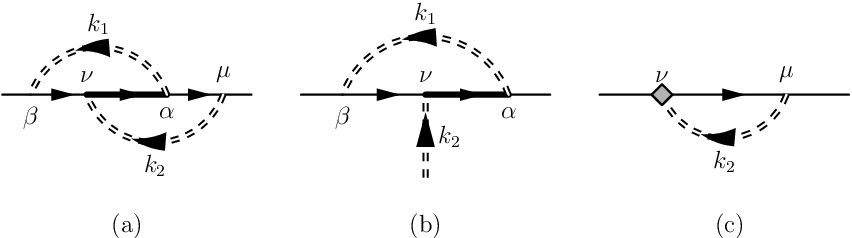,width=0.8\textwidth}
\end{center}
\caption{Two-loop diagram and diagrams corresponding to $k_1$ and
subsequent $k_2$ integrations. The diamond-shaped vertex
corresponds to the result of the $k_1$
integration.\label{Ren:TwoLoopFicDia}}
\end{figure}

   As an example consider the diagram of
Fig.~\ref{Ren:TwoLoopDia}.
   Ignoring constant factors, one can show that in a calculation up to
order ${\cal O}(q^6)$ the nucleon mass only receives contributions
from
\begin{equation}\label{Ren:Epsilon1}
    \gamma^\mu
    (\psl-m)\gamma^\alpha(\psl+m)\gamma^\nu(\psl-m)\gamma^\beta
    \iint \frac{d^{n+2}k_1 \, d^{n+2}k_2}{(2\pi)^{2n+4}} \frac{ g^{\alpha\beta}g^{\mu\nu} }{ABCDE},
\end{equation}
where the denominators are given in Eq.~(\ref{Ren:Denom}).
   One would also obtain the expression of Eq.~(\ref{Ren:Epsilon1}) if one considered a
diagram with fictitious particles as shown in
Fig.~\ref{Ren:TwoLoopFicDia}~(a), with Feynman rules given by
\begin{displaymath}
\begin{array}{cc}
\parbox{4cm}{\epsfig{file=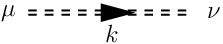,width=3cm}} &
\displaystyle{\frac{g^{\mu\nu}}{k^2-M^2+i0^+}\,,}
   \\
   \\
\parbox{4cm}{\epsfig{file=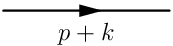,width=3cm}}& \displaystyle{\frac{\psl-m}{k^2+2p\cdot k+i0^+}\,,}
   \\
   \\
\parbox{4cm}{\epsfig{file=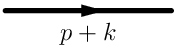,width=3cm}}& \displaystyle{\frac{\psl+m}{k^2+2p\cdot k+i0^+}\,,}
   \\
   \\
\parbox{4cm}{\epsfig{file=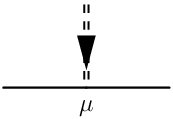,width=3cm}}& \displaystyle{\gamma_\mu\,,}
   \\
   \\
\parbox{4cm}{\epsfig{file=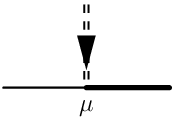,width=3cm}}& \displaystyle{\gamma_\mu\,.}
\end{array}
\end{displaymath}
   The subintegral corresponding to performing the $k_1$ integration first
is shown in Fig.~\ref{Ren:TwoLoopFicDia}~(b).
   With the Feynman rules above it is proportional to
\begin{equation}\label{Ren:Epsilon2}
 (n-3) \left(4m^2\gamma_\nu-4m p_\nu\right),
\end{equation}
so that we can identify $\varphi_1(\epsilon)=n-3=1+2\epsilon$.
   The subsequent $k_2$ integration corresponds to the diagram of
Fig.~\ref{Ren:TwoLoopFicDia}~(c), where the diamond-shaped vertex
is given by the result of the $k_1$ integration.
   One finds that the term proportional to $p_\nu$ only contributes to
higher orders and can be ignored.
   The remaining expression is proportional to
\begin{equation}\label{Ren:Epsilon3}
 (n-3)\gamma_\mu (\psl-m)\gamma_\nu g^{\mu\nu}=-2m(n-3)(n-1),
\end{equation}
and therefore $\varphi_2(\epsilon)=n-1=3+2\epsilon$.
   On the other hand, considering the $k_2$ integration first
leads to an analogous analysis with the results
$\tilde\varphi_2(\epsilon)=n-3=1+2\epsilon=\varphi_1(\epsilon)$
and
$\tilde\varphi_1(\epsilon)=n-1=3+2\epsilon=\varphi_2(\epsilon)$.

   In the cases where one cannot identify the individual contributions
to $\phi(\epsilon)$ from the integrations of $k_1$ and $k_2$,
respectively (this happens for example for tensor integrals of the
type $k_1^\mu k_2^\nu$), one has to perform the dimensional
counting analysis before reducing the tensor integrals.

\section{Application: Nucleon mass to order ${\cal
O}(q^6)$}\label{sec:NucMass}

   As an application we consider the nucleon mass up to and
including order ${\cal O}(q^6)$.\footnote{Here we consider perfect
isospin symmetry.}
   The result for the chiral expansion obtained from this
calculation has been published in Ref.~\cite{Schindler:2006ha}.
   Here, we present more details of the calculations.

\subsection{Lagrangian and power counting}
   The effective Lagrangian is given by the sum of a
purely mesonic and a one-nucleon part,
   \begin{equation}\label{Lag}
    {\cal L}_{\rm eff} = {\cal L}_{2} + {\cal L}_{4} + \cdots + {\cal L}^{(1)}_{{\pi}N} +
{\cal L}^{(2)}_{{\pi}N} + {\cal L}^{(3)}_{{\pi}N} + {\cal
L}^{(4)}_{{\pi}N} + {\cal L}^{(5)}_{{\pi}N} + {\cal
L}^{(6)}_{{\pi}N} + \cdots.
\end{equation}
   The purely mesonic Lagrangian at order ${\cal O}(q^2)$ is given
in Ref.~\cite{Gasser:1983yg}.
   Reference \cite{Gasser:1987rb} contains
the mesonic Lagrangian at order ${\cal O}(q^4)$ as well as the
lowest-order nucleonic Lagrangian.
   We use the conventions of Ref.~\cite{Becher:1999he} for the
Lagrangian at order ${\cal O}(q^2)$ and of
Ref.~\cite{Fettes:2000gb} for the Lagrangians at order ${\cal
O}(q^3)$ and ${\cal O}(q^4)$.
   While the complete Lagrangians at order ${\cal O}(q^5)$ and ${\cal O}(q^6)$ have
not yet been constructed, up to the order we are considering,
vertices from these two Lagrangians only appear as contact terms.
   The light quark masses are proportional to the square of the
pion mass, and only analytic expressions containing the quark
masses appear in the effective Lagrangian.
   Therefore the nucleon mass does not receive any contributions
from the Lagrangian at order ${\cal O}(q^5)$ in our calculation.
   The contributions from the Lagrangian at order ${\cal O}(q^6)$
are of the form $\hat{g}_1 M^6$, where $\hat{g}_1$ denotes a
linear combination of low-energy coupling constants (LECs) from
${\cal L}^{(6)}_{{\pi}N}$.

   The bare Lagrangians are decomposed into renormalized and
counterterm parts. Here we only show explicit results obtained
from the renormalized Lagrangians, i.e., all appearing coupling
constants are renormalized coupling constants. The renormalization
procedure can then be viewed as simply replacing loop integrals by
their infrared singular parts.

   We use the following standard power counting
\cite{Weinberg:1991um,Ecker:1994gg}:
   Each loop integration in $n$
dimensions is counted as $q^n$, a pion propagator as $q^{-2}$, a
nucleon propagator as $q^{-1}$ and vertices derived from ${\cal
L}_{i}$ and ${\cal L}^{(j)}_{{\pi}N}$ as $q^i$ and $q^j$,
respectively.

\subsection{Inclusion of contact interaction insertions}
   To simplify the calculation we include the self-energy
contributions from contact term diagrams in the nucleon propagator
\cite{Becher:1999he}.
   The advantage of this choice is that all self-energy diagrams with
contact interaction insertions in the propagator are summed up
automatically.

   In terms of the nucleon mass in the chiral limit $m$ the full nucleon propagator can be written as
\begin{equation}\label{Mass:prop}
    S_N(\psl)=\frac{1}{\psl-m-\Sigma(\psl,m)+i0^+},
\end{equation}
where $-i\Sigma(\psl,m)$ is the sum of all one-particle
irreducible self-energy diagrams.
   The physical nucleon mass $m_N$ is determined by the solution to the equation
\begin{equation}\label{Mass:invprop}
    \left.S_N^{-1}\right|_{\psl=m_N}=\left.[\psl-m-\Sigma(\psl,m)]\right|_{\psl=m_N}=0.
\end{equation}
   The self-energy receives contributions from contact terms
as well as from loop diagrams,
\begin{equation}\label{Mass:SEcon}
   \Sigma(\psl,m)=\Sigma_{c}+\Sigma_{loop}(\psl,m).
\end{equation}
   Due to the form of the BChPT Lagrangian used here,
$\Sigma_{c}$ for the nucleon is independent of $\psl$.
   Inserting Eq.~(\ref{Mass:SEcon}) into Eq.~(\ref{Mass:invprop}) one finds
\begin{equation}\label{Mass:invprop2}
    \left.[\psl-m-\Sigma_{c}-\Sigma_{loop}(\psl,m)]\right|_{\psl=m_N}=0.
\end{equation}
   In a loop expansion Eq.~(\ref{Mass:invprop2})
has the perturbative solution
\begin{equation}\label{Mass:mN}
    m_N=m+\Sigma_{c}+{\cal O}({\hbar}).
\end{equation}
   In the above the propagator which is used in the calculation of the
self-energy diagrams has been chosen to be
\begin{equation}\label{Mass:bareprop}
    S_N(\psl)=\frac{1}{\psl-m+i0^+}\,.
\end{equation}
   However, one can also choose this propagator to be
\begin{equation}\label{Mass:barepropnew}
    \ti{S}_N(\psl)=\frac{1}{\psl-m-\Sigma_{c}+i0^+}\,.
\end{equation}
   This corresponds to including in the free Lagrangian those terms
bilinear in $\bar \Psi, \Psi$ which generate the contact term diagrams in
the self-energy contribution.
   The advantage of this choice is that all self-energy diagrams
with contact interaction insertions in the propagator are summed
up automatically.
   With this choice of the propagator the self-energy is
now given by the sum of {\it loop} diagrams only, i.\,e.
\begin{equation}\label{Mass:SEnew}
    \Sigma(\psl,m)\rightarrow\st_{loop}(\psl,\mt)\,,
\end{equation}
where
\begin{equation}\label{Mass:mt}
    \mt=m+\Sigma_{c}\,.
\end{equation}
   As an additional benefit, when working to
two-loop accuracy, one can set $\psl=\mt$ in the expression of
two-loop diagrams, since corrections are at least of order ${\cal
O}({\hbar^3})$.
   To obtain $m_N$ one has to solve the equation
\begin{equation}\label{Mass:invpropnew}
    \ti{S}_N^{-1}(m_N)=\left.[\psl-\mt-\st_{loop}(\psl,\mt)]\right|_{\psl=m_N}=0\,.
\end{equation}
   Inserting the loop expansion for $\st_{loop}(\psl,\mt)$,
\begin{equation}\label{Mass:loopexp}
    \st_{loop}(\psl,\mt)=\hbar\st_{loop}^{(1)}(\psl,\mt)
    +\hbar^2\st_{loop}^{(2)}(\psl,\mt)+\cdots
    \,,
\end{equation}
using the ansatz
\begin{equation}\label{Mass:ansatz}
    m_N=\mt+\hbar{\Delta}m_1+\hbar^2{\Delta}m_2+\cdots
\end{equation}
and expanding around $\mt$ we obtain up to the two-loop level
\begin{align}\label{Mass:mN2loop}
    0&=\mt+\hbar{\Delta}m_1+\hbar^2{\Delta}m_2-\mt-
    \hbar\st_{loop}^{(1)}(\mt+\hbar{\Delta}m_1,\mt)
    -\hbar^2\st_{loop}^{(2)}(\mt,\mt)  \notag\\
    &=\hbar\left[{\Delta}m_1-\st_{loop}^{(1)}(\mt,\mt)\right]
    +\hbar^2\left[{\Delta}m_2-{\Delta}m_1\,\st_{loop}^{(1)\prime}(\mt,\mt)
    -\st_{loop}^{(2)}(\mt,\mt)\right],
\end{align}
where $\st_{loop}^{(1)\prime}(\psl,\mt)$ denotes the
derivative of $\st_{loop}^{(1)}(\psl,\mt)$ with respect to
$\psl$.
   The solutions for ${\Delta}m_1$ and ${\Delta}m_2$ are given by
\begin{align}\label{Mass:dm1}
  {\Delta}m_1 &= \st_{loop}^{(1)}(\mt,\mt), \\
  {\Delta}m_2 &= \st_{loop}^{(1)}(\mt,\mt)\st_{loop}^{(1)\prime}(\mt,\mt)
  +\st_{loop}^{(2)}(\mt,\mt)\,. \label{Mass:dm2}
\end{align}
   To obtain the nucleon mass up to chiral order ${\cal O}(q^6)$
one needs to determine $\Sigma_{c}$, ${\Delta}m_1$ and
${\Delta}m_2$ up to that order.
In the following we will not directly evaluate ${\Delta}m_2$ as given in Eq.~(\ref{Mass:dm2}) since we are only intested in the combination $\hbar{\Delta}m_1+\hbar^2{\Delta}m_2$ with $\hbar=1$. 
Instead, as indicated in the first line of Eq.~(\ref{Mass:mN2loop}), we will use the result for ${\Delta}m_1$ to determine $\st_{loop}^{(1)}(\mt+\hbar{\Delta}m_1,\mt)$ directly (see Eq.~(\ref{Mass:OneLoopRenorm}) below).

   In principle, the nucleon propagator is a $2\times 2$-matrix
in isospin space.
   For arbitrary values of the up and down quark masses the
propagator is a diagonal matrix; since however in this work the
isospin-symmetric case $m_u=m_d$ is considered, the masses of proton
and neutron are identical and the propagator is proportional to
the unit matrix.

\subsection{Contact terms}
   The contributions to the nucleon mass from contact interactions
are given by
\begin{align}\label{Mass:Contact}
    \delta m_c &= -4 c_1 M^2  - (16e_{38} + 2e_{115} + 2e_{116} ) M^4
+\hat{g}_1 M^6\,\notag\\
    &= -4 c_1 M^2  - \hat{e}_1 M^4 + \hat{g}_1 M^6\,,
\end{align}
where $M^2$ is the lowest-order expression for the square of the
pion mass. We use the notation
\begin{equation}\label{Mass:e1hat}
\hat{e}_1=16e_{38}+2e_{115}+2e_{116}
\end{equation}
and $\hat{g}_1$ denotes a linear combination of LECs from the
Lagrangian at order ${\cal O}(q^6)$.

\subsection{One-loop diagrams}
\begin{figure}
\begin{center}
\epsfig{file=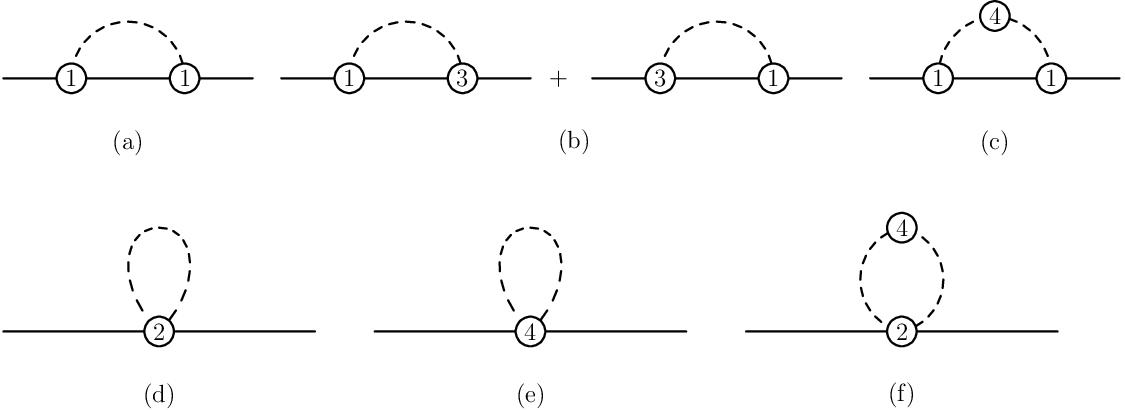,width=\textwidth}
\end{center}
\caption{One-loop diagrams contributing to the nucleon self-energy
up to order ${\cal O}(q^6)$.\label{Mass:OneLoopDia}}
\end{figure}

   The one-loop diagrams contributing to the nucleon mass
up to order ${\cal O}(q^6)$ are shown in
Fig.~(\ref{Mass:OneLoopDia}).
   Diagrams (a) and (d) are of order ${\cal O}(q^3)$ and ${\cal
O}(q^4)$, respectively, and have been determined in
Ref.~\cite{Becher:1999he}.
   Diagrams (b) and (c) are of order ${\cal O}(q^5)$, while the power
counting gives $D=6$ for diagrams (e) and (f).

   Using dimensional regularization the unrenormalized results for
the one-loop diagrams up to order ${\cal O}(q^6)$ read
\begin{align}\label{Mass:OneLoopUnrenorm}
  \Sigma_{1(a)} &= -\frac{3\texttt{g}_A^2}{4F^2}\left[(\psl+\mt)H_N+(\psl+\mt)M^2H_{{\pi}N}(p^2)
+(p^2-\mt^2){\psl}H_{{\pi}N}^{(p)}(p^2)\right],\notag\\
  \Sigma_{1(b)} &= -\frac{3\texttt{g}_A}{F^2}(2d_{16}-d_{18})M^2
\left[(\psl+\mt)H_N+(\psl+\mt)M^2H_{{\pi}N}(p^2)\right. \notag\\
  & \quad \left. +(p^2-\mt^2){\psl}H_{{\pi}N}^{(p)}(p^2)\right],\notag\\
  \Sigma_{1(c)} &= -\frac{3\texttt{g}_A^2}{2F^4}\,(\psl+\mt) M^2 \left\{l_3 M^2 \left[H_{\pi\pi}
-(p^2-\mt^2)H_{\pi\pi N}(p^2)-\psl(\psl+\mt)H_{\pi\pi N}^{(p)}(p^2) \right] \right. \notag\\
  & \left. -l_4 \left[ H_\pi -(p^2-\mt^2)H_{\pi N}(p^2)-\psl(\psl+\mt)H_{\pi N}^{(p)} \right] \right\} ,\notag\\
  \Sigma_{1(d)} &= \frac{3}{F^2}\left[(2c_1-c_3)M^2 H_\pi
- c_2 \frac{p^2}{m^2}\,H_\pi^{(00)} \right],\notag\\
  \Sigma_{1(e)} &= -\frac{12}{F^2}\,\left\{\left[2(e_{14}+e_{19})-e_{36}-4e_{38}\right]
M^4 H_\pi +2\left[e_{15}+e_{20}+e_{35}\right]\frac{p^2}{m^2}M^2 H_\pi^{(00)}\right.\notag\\
  &\quad \left. +6e_{16}\frac{p^4}{m^4}\, H_\pi^{(0000)}\right\}, \notag\\
\Sigma_{1(f)} &= 6\,\frac{M^4}{F^4}\left\{2 c_1 \left[ l_3 M^2 H_{\pi\pi}-l_4 H_\pi \right]  -\frac{c_2}{n}\frac{p^2}{m^2}\left[(l_3-l_4)H_\pi+l_3 M^2 H_{\pi\pi} \right]\right.\notag\\
  &\quad \left. -c_3 \left[ (l_3-l_4)H_\pi-l_3 M^2 H_{\pi\pi} \right] \right\}.
\end{align}
   The integrals $H_\pi, H_{{\pi}N}(p^2),\ldots$ are given in
App.~\ref{App:IntDef}.
      Various combinations of fourth-order baryonic LECs appear through the
vertex in diagram (e).
   To simplify the notation we use $\hat{e}_1$ as defined in Eq.~(\ref{Mass:e1hat}) and
\begin{align}\label{Mass:eComb}
  \hat{e}_2 &= 2e_{14}+2e_{19}-e_{36}-4e_{38} \,,  \\
  \hat{e}_3 &= e_{15}+e_{20}+e_{35} \notag
\end{align}
for these combinations in the following.

   To determine the contribution of these diagrams to the nucleon
mass we evaluate the expressions of
Eq.~(\ref{Mass:OneLoopUnrenorm}) at $p^2=m_N^2$ between on-shell
spinors.
To the order we are working, we can use $m_N=\mt+\hbar{\Delta}m_1$ and thus obtain $\st_{loop}^{(1)}(\mt+\hbar{\Delta}m_1,\mt)$ in Eq.~(\ref{Mass:mN2loop}).
   Further, we renormalize the one-loop integrals by replacing them with the corresponding
infrared singular parts.
   The infrared renormalized expressions for the mass contributions, denoted by a superscript $r$,
up to order $M^6$ are given by
\begin{align}\allowdisplaybreaks\label{Mass:OneLoopRenorm}
  \delta m_{1(a)}^{\, r} &= -\frac{3 \texttt{g}_A^2}{32 \pi F^2}\,M^3
- \frac{3 \texttt{g}_A^2}{64 \pi^2 F^2 m} \left[2 \logM +1\right]\,M^4 \notag\\
   & \quad + \frac{3\texttt{g}_A^2}{1024 \pi^3 F^4 m^2} \left[4 \pi^2 F^2+3 \texttt{g}_A^2
m^2+9 \texttt{g}_A^2 m^2 \logM \right]\,M^5 \notag\\
   & \quad -\frac{\texttt{g}_A^2}{2048 \pi^4 F^4 m^3} \Bigg[ 27 \pi^2 \texttt{g}_A^2 m^2+384 \pi^2 c_1 F^2 m
-16 \pi^2 F^2-9m^2(\texttt{g}_A^2-c_2 m)  \notag\\
   & \quad  +3m\left[-15\texttt{g}_A^2 m +16c_1 (3m^2+16\pi^2F^2)+3m^2(c_2-8c_3)\right]\logM \notag\\
   & \quad -54 m^2\left(\texttt{g}_A^2-8c_1m+c_2m+4c_3m\right)\logMsq\Bigg]\,M^6 \,, \notag \\
   \delta m_{1(b)}^{\, r} &= -\frac{3 \texttt{g}_A}{8\pi F^2 }\,\left(2 d_{16}-d_{18}\right)M^5
-\frac{3 \texttt{g}_A}{16\pi^2F^2 m}\,
\left(2d_{16}-d_{18}\right) \left[2 \logM +1\right]M^6  \,, \notag\\
   \delta m^{\, r}_{1(c)} &= -\frac{3 \texttt{g}_A^2}{32\pi F^4}\,(3l_3-2l_4)M^5-\frac{3 \texttt{g}_A^2}{32\pi^2 F^4m}
\left[3l_3-l_4+2(2l_3-l_4)\logM\right] M^6 \,, \notag\\
   \delta m^{\, r}_{1(d)} &= \frac{3}{128\pi^2F^2}
\left[c_2+\logM\left(32c_1-4 c_2-16c_3\right)\right] M^4
+\frac{3c_1 c_2 }{16\pi^2 F^2m}\left[4\logM -1\right]M^6 ,\notag\\
   \delta m^{\, r}_{1(e)} &= \frac{M^6}{32\pi^2F^2}\,
\left[6\hat e_{3}+5e_{16} \right] -\frac{3M^6}{8\pi^2 F^2} \ln\frac{M}{\mu}
\left[4\hat e_{2}+2\hat e_{3} +e_{16}\right]\,,\notag\\
\delta m^{\, r}_{1(f)} &= \frac{3}{64\pi^2 F^4}\left[ 16c_1 l_3 -c_2 l_4 -8c_3l_3 \right]M^6 + \frac{3}{16\pi^2F^4}\left[ 8c_1(l_3-l_4) \right.\notag\\
   & \quad \left.-(c_2+4c_3)(2l_3-l_4) \right]M^6 \logM.
\end{align}
The scale dependence of the renormalized low-energy constants is
governed by
\begin{equation}\label{RenormCoup}
l_{i,0}=l_i(\mu)+\gamma_i \lambda,\quad d_{i,0}=d_i(\mu)+\frac{\delta_i}{F^2}\lambda,
\quad e_{i,0}=e_i(\mu) + \frac{\varepsilon_i}{m F^2}\lambda,
\end{equation}
where the subscript $0$ denotes bare quantities and
$$
\lambda = \frac{\mu^{n-4}}{16\pi^2}\left\{
\frac{1}{n-4}-\frac{1}{2} \left[ \ln (4\pi)
+\Gamma'(1)+1\right]\right\}.
$$
   The coefficients $\gamma_i$ are given by \cite{Gasser:1983yg}
\begin{equation}\label{gamma}
\gamma_3=-\frac{1}{2},\quad \gamma_4=2,
\end{equation}
while the $\delta_i$ can be taken from
Ref.~\cite{Ecker:1995rk},\footnote{Note that the numbering of
terms in the Lagrangian used here differs from
Ref.~\cite{Ecker:1995rk}.}
\begin{equation}\label{delta}
\delta_{16}=\frac{1}{2}\texttt{g}_A+\texttt{g}_A^3,\quad \delta_{18}=0.
\end{equation}
The LECs $\hat e_2,\hat e_3, e_{16}$ appear through the vertex shown in Fig.~\ref{Mass:OneLoopDia}~(e), which also gives the contact term contribution at order ${\cal O}(q^4)$ to $\pi N$ scattering as analyzed in Ref.~\cite{Becher:2001hv}.
We can therefore relate the combinations of LECs used here to the ones in Ref.~\cite{Becher:2001hv} (here denoted by a superscript $BL$), resulting in
$$
\hat e_1 = -e_1^{BL}, \quad \hat e_2 = \frac{1}{8}e_3^{BL}, \quad \hat e_3 = \frac{1}{16}e_4^{BL}, \quad e_{16}=\frac{1}{16}e_6^{BL}.
$$
Using the expressions for the renormalized couplings given in App.~E of Ref.~\cite{Becher:2001hv} we find
\begin{align}\label{varepsilon}
\hat \varepsilon_1 &= -\frac{3}{2}\texttt{g}_A+\frac{3}{2}(8c_1-c_2-4c_3)m,\notag\\
\hat \varepsilon_2 &= -\frac{1}{8}\left( 1+3\texttt{g}_A^2+\frac{22}{3}\texttt{g}_A^4+8c_1m+c_2 m-4c_3m \right),\notag\\
\hat \varepsilon_3 &= \frac{1}{16}\left(10+12\texttt{g}_A^2+\frac{52}{3}\texttt{g}_A^4+8c_2m \right) ,\notag\\
\varepsilon_{16} &= -\frac{1}{16}\left(12+8\texttt{g}_A^2+8\texttt{g}_A^4 \right).
\end{align}

\subsection{Two-loop diagrams}
\begin{figure}
\begin{center}
\epsfig{file=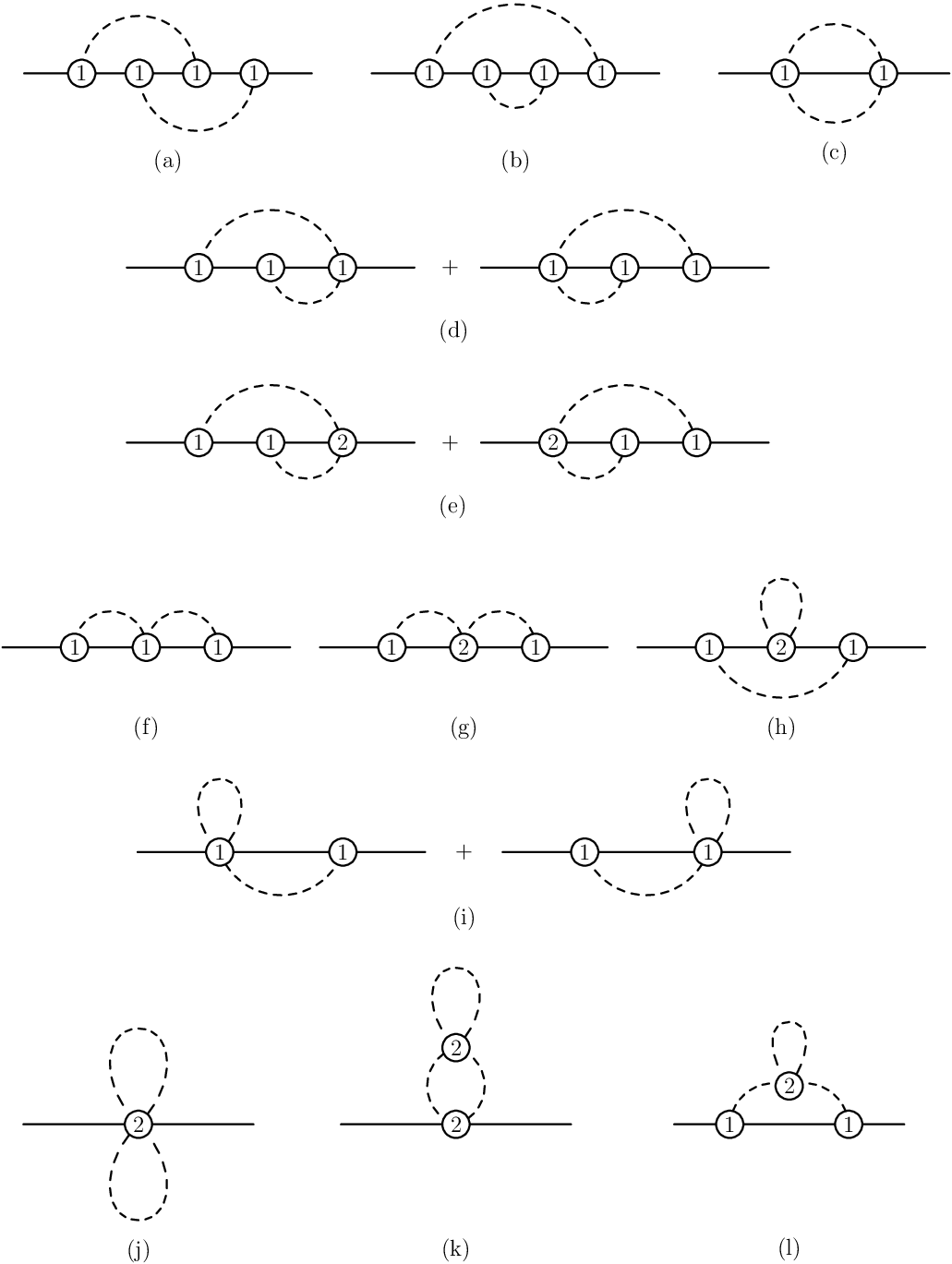,height=0.9\textheight}
\end{center}
\caption{Two-loop diagrams contributing to the nucleon self-energy
up to order ${\cal O}(q^6)$.\label{Mass:TwoLoopDia}}
\end{figure}

   The two-loop diagrams relevant for a calculation of the nucleon
self-energy up to order ${\cal O}(q^6)$ are shown in
Fig.~\ref{Mass:TwoLoopDia}.
   According to the power counting there are further diagrams
at the given order.
   An example would be diagram \ref{Mass:TwoLoopDia}~(c) with one first-order vertex replaced by a
second-order one.
   As a result of our calculation we find that these diagrams give
vanishing contributions to the nucleon mass up to the order we are
considering.

   We again employ dimensional regularization.
   The unrenormalized expressions for the mass contributions
of the diagrams of Fig.~\ref{Mass:TwoLoopDia} up to order ${\cal
O}(q^6)$ can be reduced to
\begin{align}\label{Mass:TwoLoopNonRen}\allowdisplaybreaks
   \delta m_{2(a)} &=  -\frac{6 \texttt{g}_A^4}{F^4}\,\pi^2 m^3 (n-1)(n-3)
H_2(1,1,1,1,1|n+2)  \,, \notag \\
   \delta m_{2(b)} &=  \frac{9\texttt{g}_A^4}{F^4}\,m \pi^2 (n-1)\left\{2m^2 H_2(1,1,1,0,2|n+2)
-M^2H_2(1,2,1,0,1|n+2)\right.  \notag\\
   & \quad  -4m^2M^2\left[H_2(1,2,2,0,1|n+2) +H_2(1,1,2,0,2|n+2)\right]  \notag\\
   & \quad \left. -32\pi^2m^2\left[2H_2(1,2,1,0,3|n+4)+H_2(1,2,2,0,2|n+4)\right]\right\} \,, \notag \\
   \delta m_{2(c)} &= \frac{1}{2}\frac{3}{F^4}\, m
\left\{ \frac{1}{2}\,H_2(1,1,0,0,0|n) -M^2H_2(1,1,0,0,1|n)
-H_2(1,0,0,0,1|n) \right. \notag\\
   & \quad + 8\pi^2(n-1)H_2(1,1,0,0,2|n+2)-16\pi^2(M^2-2m^2)H_2(1,2,0,0,2|n+2)\bigg\} \,, \notag \\
   \delta m_{2(d)} &= \frac{24 \texttt{g}_A^2}{F^4}\,\left[32\pi^2 m^3(n-1)
\left[2H_2(1,2,1,0,3|n+4)+H_2(1,2,2,0,2|n+4)\right]\right.  \notag\\
   & \quad \left.   +\pi^2 m M^2
\left[(n-1)H_2(1,2,1,0,1|n+2)+H_2(1,2,0,0,2|n+2)\right]\right] \,, \notag \\
   \delta m_{2(e)} &=  - \frac{48 \texttt{g}_A^2}{F^4}\,\pi^2 m^2 (n-1)
\left\{c_3 \! \left[H_2(1,1,0,0,2|n+2) \! -64\pi^2m^2 H_2(2,1,1,0,3|n+4)\right]\right.  \notag\\
   & \quad +c_4\left[2H_2(1,1,0,0,2|n+2)-2M^2H_2(1,2,1,0,1|n+2) \right.\notag\\
   & \quad \left.\left. -64\pi^2m^2H_2(2,1,1,0,3|n+4)\right]\right\} \,, \notag \\
   \delta m_{2(f)} &= \frac{3 \texttt{g}_A^2}{2F^4}\, m M^4H_{{\pi}N}(1,1|n)
H_{{\pi}N}(1,1|n)  \,, \notag \\
   \delta m_{2(g)} &= \frac{24 \texttt{g}_A^2}{F^4}\,\pi^2 m^2
(n-1)\left[(n-2)c_4-c_3\right]H_{{\pi}N}(1,1|n+2)H_{{\pi}N}(1,1|n+2)  \,, \notag \\
   \delta m_{2(h)} &= \frac{18 \texttt{g}_A^2}{F^4}\,\pi m^2 M^2 (n-1)
\left[ \frac{c_2}{n}+c_3-2c_1 \right] H_{{\pi}N}(1,0|n)H_{{\pi}N}(1,2|n+2)  \notag\\
   &= - \frac{72 \texttt{g}_A^2}{F^4}\,\pi^2 m^2 M^2 (n-1)
\left[ \frac{c_2}{n}+c_3-2c_1 \right]
H_{{\pi}N}(2,0|n+2)H_{{\pi}N}(1,2|n+2)  \,, \notag \\
   \delta m_{2(i)} &=  -\frac{\texttt{g}_A^2}{F^4}\, m M^2
H_{{\pi}N}(1,0|n)H_{{\pi}N}(1,1|n)  \,, \notag \\
   \delta m_{2(j)} &= \frac{1}{8} \frac{4M^2}{F^4}\,
\left[5c_1-4\frac{c_2}{n}-4c_3\right]H_{{\pi}N}(1,0|n)H_{{\pi}N}(1,0|n)  \,, \notag \\
   \delta m_{2(k)} &=  \frac{1}{4}
\frac{2}{F^4}\,\left\{3\left[\frac{c_2}{n}+c_3-2c_1 \right]M^4
H_{{\pi}N}(1,0|n) H_{{\pi}N}(2,0|n)\right.
\notag\\
   & \quad +\left.\left[7\frac{c_2}{n}+7c_3-8c_1
\right]M^2H_{{\pi}N}(1,0|n)H_{{\pi}N}(1,0|n)\right\}  \,, \notag \\
   \delta m_{2(l)} &= \frac{\texttt{g}_A^2}{4F^4}m  H_{{\pi}N}(1,0|n) \left[4H_{{\pi}N}(0,1|n) +7M^2H_{{\pi}N}(1,1|n)
+3M^4H_{{\pi}N}(2,1|n) \right] \,.
\end{align}
   The integrals $H_{{\pi}N}(a,b|n)$ and $H_2(a,b,c,d,e|n)$
are defined in App.~\ref{App:TwoLoopIntDef}.
   Here we have expressed tensor integrals in terms of scalar
integrals in higher dimensions where convenient (see
App.~\ref{App:TwoLoopIntDef} and also Ref.~\cite{Tarasov:1996br}).

   After performing the infrared renormalization as described in Sec.~\ref{Sec:TwoLoop} the
contributions to the nucleon mass up to order ${\cal O}(q^6)$ read
\begin{align}\label{Mass:TwoLoopRen}
   \delta m^{\, r}_{2(a)} &= -\frac{\texttt{g}_A^4}{512 \pi^3 F^4}\, \left[3 M^5\left(1+\logM\right)
-\frac{M^6}{48\pi m}\left(5+36\pi^2+48\logM\right)\right], \notag \\
   \delta m^{\, r}_{2(b)} &= \frac{\texttt{g}_A^4}{F^4}\left[\frac{9}{1024\pi^3}M^5\left(1+3\logM\right)
-\frac{27}{4096\pi^4 m}M^6\left(1+6\logM+4\logMsq\right)\right], \notag \\
   \delta m^{\, r}_{2(c)} &= \frac{M^6}{2048 \pi^4 F^4 m} \,, \notag \\
   \delta m^{\, r}_{2(d)} &= - \frac{\texttt{g}_A^2}{1536\pi^4 F^4 m}\,M^6 \left[1+9\pi^2-6\logM\right], \notag \\
   \delta m^{\, r}_{2(e)} &= \frac{\texttt{g}_A^2}{128\pi^2F^4}\,M^6\left[c_3-2c_4\right], \notag \\
   \delta m^{\, r}_{2(f)} &= - \frac{3 \texttt{g}_A^2}{512 \pi^2 F^4 m}\, M^6 \,, \notag \\
   \delta m^{\, r}_{2(g)} &= \frac{\texttt{g}_A^2}{128 \pi^2 F^4}\, \left[c_3-2c_4\right] M^6\,, \notag \\
   \delta m^{\, r}_{2(h)} &= - \frac{9\texttt{g}_A^2}{256\pi^4F^4}\,M^6 \left[(c_3-2c_1)
\left(\logM+3\logMsq\right)+
\frac{c_2}{16}\left(-1+\logM \right. \right. \notag\\
 & \quad \left.\left. +12\logMsq\right)\right], \notag \\
   \delta m^{\, r}_{2(i)} &=  \frac{\texttt{g}_A^2}{128\pi^3F^4}\,\logM\, M^5
+ \frac{\texttt{g}_A^2}{256\pi^4 F^4 m}\,\left(\logM+2\logMsq\right)M^6 \,, \notag \\
   \delta m^{\, r}_{2(j)} &= -\frac{M^6}{128\pi^4F^4}\,
\left[(5c_1-c_2-4c_3)\logMsq+\frac{c_2}{4}\logM\right]  , \notag \\
   \delta m^{\, r}_{2(k)} &= \frac{M^6}{512\pi^4F^4}\left[(12 c_1+c_2-6c_3) \logM +2(28c_1-5
c_2-20c_3) \logMsq\right] , \notag \\
   \delta m^{\, r}_{2(l)} &=  -\frac{17\texttt{g}_A^2}{1024\pi^3F^4}\,\logM\,M^5
- \frac{\texttt{g}_A^2}{1024\pi^4F^4 m}\left(13 \logM +
20\logMsq\right)\,M^6  \,.
\end{align}

\subsection{Results and discussion}
   Combining the contributions from the contact interactions with the one- and
two-loop results we obtain for the nucleon mass up to order ${\cal
O}(q^6)$
\begin{align}\label{Mass:Exp}
    m_N &= m +k_1 M^2 +k_2 \,M^3 +k_3 M^4 \logM
+ k_4 M^4  + k_5 M^5\logM + k_6 M^5  \notag \\
    & \qquad + k_7 M^6 \logMsq + k_8 M^6 \logM + k_9 M^6 \,.
\end{align}
   The coefficients $k_i$ are given by
\begin{align}\label{Mass:Coeff}
   k_1 &= -4 c_1 \,,\notag\\
   k_2 &= -\frac{3 \texttt{g}_A^2}{32 \pi F^2} \,,\notag\\
   k_3 &= - \frac{3}{32\pi^2F^2m}
\left(\texttt{g}_A^2-8c_1m +c_2 m +4c_3m\right)\,,\notag\\
   k_4 &= - \hat{e}_1
-\frac{3}{128\pi^2F^2m}\left(2 \texttt{g}_A^2-c_2m\right) \,,\notag\\
   k_5 &= \frac{3 \texttt{g}_A^2}{1024\pi^3 F^4}\,\left(16\texttt{g}_A^2-3\right) \,,\notag\\
   k_6 &= \frac{3 \texttt{g}_A^2}{256 \pi^3 F^4}\, \left[ \texttt{g}_A^2 + \frac{\pi^2 F^2}{m^2}
-8\pi^2(3l_3-2l_4) -\frac{32\pi^2F^2}{\texttt{g}_A}\,(2d_{16}-d_{18}) \right] \,,\notag\\
   k_7 &=  -\frac{3}{256\pi^4 F^4 m}\,\left[\texttt{g}_A^2 - 6 c_1 m + c_2m +4c_3m
  \right]  \,,\notag\\
   k_8 &= -\frac{\texttt{g}_A^4}{64\pi^4 F^4 m}-\frac{\texttt{g}_A^2}{1024\pi^4 F^4 m^2}\left[384 \pi^2 F^2 c_1
+5m+192\pi^2m(2l_3-l_4)\right] \notag\\
   & \quad -\frac{3\texttt{g}_A}{8\pi^2F^2m}\left[2d_{16}-d_{18}\right] +\frac{3}{256\pi^4F^4}\left[ 2c_1-c_3\right]
+\frac{3}{8\pi^2F^2m}\left[ 2c_1c_2-4\hat{e}_2 m-2\hat{e}_3m -e_{16}m  \right]\notag\\
   & \quad +\frac{3}{16\pi^2F^4}\left[ 8c_1 (l_3-l_4)-(c_2+4c_3)(2l_3-l_4)\right]\,,\notag\\
   k_9 &=  \hat{g}_1 - \frac{\texttt{g}_A^4}{24576\pi^4 F^4 m}\left( 49+288\pi^2 \right)
-\frac{3\texttt{g}_A}{16\pi^2 F^2 m} \left( 2d_{16}-d_{18} \right)
\notag\\
   & \quad -\frac{\texttt{g}_A^2}{1536\pi^4 F^4 m^3}\left[ m^2(1+18\pi^2)-12\pi^2 F^2 +144\pi^2 m^2\left(3l_3-l_4\right)
\right.  \notag\\
   & \quad +\left. 288\pi^2 F^2 m c_1 -24\pi^2 m^3 \left(c_3-2c_4\right)\right]
+\frac{3}{64\pi^2F^4}\left[ 8 (2c_1-c_3) l_3 - c_2 l_4\right] \notag\\
   & \quad +\frac{1}{2048\pi^4 F^4 m}\left[1-384\pi^2 F^2 c_1 c_2 +384\pi^2 F^2 m\, \hat{e}_3
+320\pi^2F^2m\, e_{16}\right]\,.
\end{align}

   In general, the expressions of the coefficients in the chiral
expansion of a physical quantity differ in various renormalization
schemes, since analytic contributions can be absorbed by
redefining LECs.
   However, this is not possible for the leading nonanalytic
terms, which therefore have to agree in all renormalization
schemes.
   Comparing our result with the HBChPT calculation of
\cite{McGovern:1998tm}, we see that the expressions for the
coefficients $k_2$, $k_3$, and $k_5$ agree as expected.
   At order ${\cal O}(q^6)$ also the coefficient $k_7$ has to be
the same in all renormalization schemes.
   Note that, while $k_6 M^5$ and $k_8 M^6 \logM$ are
nonanalytic in the quark masses, the algebraic form of the
coefficients $k_6$ and $k_8$ are renormalization scheme
\emph{dependent}.
   This is due to the different treatment of one-loop diagrams in
different renormalization schemes.
   The counterterms for one-loop subdiagrams depend on the
renormalization scheme and produce nonanalytic terms proportional
to $M^5$ and $M^6 \logM$ when used as vertices in counterterm
diagrams.
   We find that our result for $k_6$ coincides with the HBChPT
calculation of Ref.~\cite{McGovern:1998tm} except for a term
proportional to $d_{28}$, which, however, does not have a finite
contribution for manifestly Lorentz-invariant renormalization
schemes \cite{Fettes:1998ud}.
   Therefore, at order ${\cal O}(q^5)$ the chiral expansion of
the IR renormalized result reproduces the HBChPT result.

   The result for the nucleon mass should be scale-independent at
each order in the chiral expansion, and showing this scale
independence serves as a check of our results. The terms up to and
including order ${\cal O}(M^4)$ have been discussed previously
\cite{Becher:2001hv}. Using the expressions for the scale
dependence of the renormalized couplings of
Eqs.~(\ref{RenormCoup}), (\ref{gamma}) and (\ref{delta}) we see
that the contribution at order ${\cal O}(M^5)$ is independent of
$\mu$ as required. At order ${\cal O}(M^6)$, the scale dependence
of the LEC $\hat g_1$ is not known which prevents a complete
analysis at this order. However, $\hat g_1$ does not contribute to
terms proportional to $\ln M \ln\mu$ since it only appears in the
analytic expression at ${\cal O}(M^6)$. We can therefore analyze
the terms proportional to $\ln M \ln\mu$ which must vanish if our
result is to be scale-independent. This is the case, which can be
shown using the expressions of
Eqs.~(\ref{RenormCoup})-(\ref{varepsilon}).

   The numerical contributions from higher-order terms cannot be
calculated so far, since most expressions in
Eq.~(\ref{Mass:Coeff}) contain LECs which are not reliably known
in IR renormalization.
   In order to get an estimate of these contributions we consider
several terms for which the LECs have previously been determined.
   The coefficient $k_5$ is free of higher-order LECs
and is given in terms of the axial-vector coupling constant
$\texttt{g}_A$ and the pion decay constant $F$.
   While the values for both $\texttt{g}_A$ and $F$ should be taken in the chiral
limit, we evaluate $k_5$ using the physical values
$g_A=1.2695(29)$ \cite{Yao:2006px} and $F_\pi=92.42(26)$ MeV.
   Setting $\mu=m_N$, $m_N=(m_p+m_n)/2=938.92$ MeV, and $M=M_{\pi^+}=139.57$ MeV
we obtain $k_5 M^5 \ln(M/m_N) = -4.8$ MeV.
   This amounts to approximately $31$\% of the leading nonanalytic contribution at one-loop
order, $k_2 M^3$.
   The mesonic LECs appearing in $k_6$ can be found in
Ref.~\cite{Gasser:1983yg} and are given by $l_3(m_N)=1.4 \times
10^{-3}$ and $l_4(m_N)=3.7 \times 10^{-3}$ at the scale $\mu=m_N$.
   The parameter $d_{18}$ can be related to the Goldberger-Treiman
discrepancy \cite{Becher:2001hv} and is given by
$d_{18}=-0.80\,\mbox{GeV}^{-2}$.
   The LEC $d_{16}$, however, is not as reliably determined.
   In order to estimate the magnitude of the contribution stemming from $k_6$
we use the central value from the reaction $\pi N \rightarrow
\pi\pi N$, $d_{16}(m_N)=-1.93\,\mbox{GeV}^{-2}$
\cite{Fettes:1999wp,Beane:2004ks}.
   It should be noted that the calculation of
Ref.~\cite{Fettes:1999wp} was performed in HBChPT, and employing
the obtained value for $d_{16}$ in an infrared renormalized
expression therefore only gives an estimate of the size of the
corresponding term.
   The resulting contribution is $k_6 M^5=3.7$ MeV and cancels
large parts of the nonanalytic term $k_5 M^5 \ln(M/m_N)$.
   In Ref.~\cite{McGovern:2006fm} the parameter $d_{16}$ has been
determined by a fit to lattice data.
   At the scale $\mu=m_N$ it is given by $d_{16}(m_N)=4.11\
\mbox{GeV}^{-2}$, which does not agree with the result from the
reaction $\pi N \rightarrow \pi\pi N$.
   With this value of $d_{16}$ we find $k_6 M^5=-7.6$ MeV.
   The LECs appearing in $k_7$ have been determined in
Ref.~\cite{Becher:2001hv}, and we obtain $k_7 M^6
\ln^2(M/m_N)=0.3$ MeV.

   The terms $k_8$ and $k_9$ contain LECs from the fourth order
Lagrangian $\mathcal{L}_{{\pi}N}^{(4)}$ which have not been
determined.
   We try to get a very rough estimate of the size of these
contributions by assuming that all these LECs as well as
$\hat{g}_1$ are of natural size, that means $e_i\sim 1\,
\mbox{GeV}^{-3}$ and $\hat{g}_1\sim 1\, \mbox{GeV}^{-5}$.
   We choose the $d_{16}$ value from $\pi N \rightarrow
\pi\pi N$ and use the above values for the other LECs.
   Setting all appearing $e_i=0\,\mbox{GeV}^{-3}$ gives a
contribution $k_8 M^6 \ln(M/m_N) \approx 10^{-2}\,\mbox{MeV}$.
   The choice $e_i=5\,\mbox{GeV}^{-3}$ results in $k_8 M^6 \ln(M/m_N) \approx
0.9\,\mbox{MeV}$, while $e_i=-5\,\mbox{GeV}^{-3}$ gives $k_8 M^6
\ln(M/m_N) \approx -0.9\,\mbox{MeV}$.
   A similar analysis for the term $k_9 M^6$ gives $k_9 M^6\approx
-2.8\,\mbox{MeV}$ for all $e_i=0\,\mbox{GeV}^{-3}$ and
$\hat{g}_1=0\,\mbox{GeV}^{-5}$, while setting
$e_i=5\,\mbox{GeV}^{-3}$, $\hat{g}_1=5\,\mbox{GeV}^{-5}$ and
$e_i=-5\,\mbox{GeV}^{-3}$, $\hat{g}_1=-5\,\mbox{GeV}^{-5}$ results
in $k_9 M^6\approx -2.5\,\mbox{MeV}$ and $k_9
M^6\approx-3.2\,\mbox{MeV}$,
 respectively.
   One should note, however, that the numbers obtained here are only
very rough estimates.
   Choosing $e_{14}=e_{15}=5\,\mbox{GeV}^{-3}$ and
$e_{16}=e_{19}=e_{20}=e_{35}=e_{36}=e_{38}=1\,\mbox{GeV}^{-3}$,
$\hat{g}_1=1\,\mbox{GeV}^{-5}$ leads to large cancelations between
the terms $k_8 M^6 \ln(M/m_N)$ and $k_9 M^6$, resulting in a
complete contribution at order ${\cal O}(M^6)$ of about
$0.3\,\mbox{MeV}$.
   As a check we also use the value of $d_{16}$ as obtained in
Ref.~\cite{McGovern:2006fm}, which results in contributions from
$k_8$ that are about a factor 10 larger, while the dependence of
$k_9$ on $d_{16}$ is much less pronounced.
   Clearly a more reliable determination of the higher-order LECs
is desirable.

\begin{figure}
\begin{center}
\epsfig{file=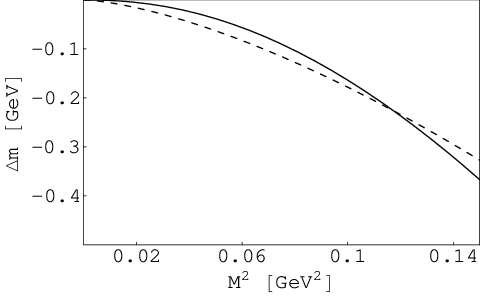,width=0.5\textwidth}
\end{center}
\caption{Pion mass dependence of the term $k_5 M^5 \ln(M/m_N)$
(solid line) for $M<400\,\mbox{MeV}$. For comparison also the term
$k_2 M^3$ (dashed line) is shown.\label{Mass:k2k5Low}}
\end{figure}

\begin{figure}
\begin{center}
\epsfig{file=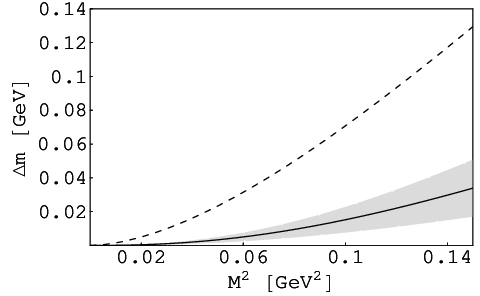,width=0.5\textwidth}
\end{center}
\caption{Pion mass dependence of the term $k_7 M^6 \ln^2(M/m_N)$
(solid line) for $M<400\,\mbox{MeV}$. The shaded band corresponds
to relative errors of $20\%$ in the LECs. For comparison also the
term $k_3 M^4 \ln(M/m_N)$ (dashed line) is
shown.\label{Mass:k3k7Low}}
\end{figure}

   Chiral expansions like Eq.~(\ref{Mass:Exp}) play an important
role in the extrapolation of lattice QCD results to physical quark
masses, and the nucleon mass is an example that has been studied
in detail (see, e.g.,
Refs.~\cite{Procura:2003ig,Bernard:2003rp,Leinweber:2003dg,Procura:2006bj,McGovern:2006fm}).
   In Ref.~\cite{Procura:2003ig} such an extrapolation was
performed for the nucleon mass up to order ${\cal O}(q^4)$, while
Ref.~\cite{McGovern:2006fm} includes an analysis of the
fifth-order terms.
   It was shown, as had also been argued in
Ref.~\cite{Beane:2004ks}, that the terms at order ${\cal O}(q^5)$
play an important role in the chiral extrapolation.
   As an illustration we consider the leading nonanalytic term at
this order, $k_5 M^5 \ln(M/m_N)$.
   Its dependence on the pion mass is shown in
Fig.~\ref{Mass:k2k5Low} for pion masses below $400\,\mbox{MeV}$,
which is considered a region where chiral extrapolations are valid
(see, e.g., Refs.~\cite{Meissner:2005ba,Djukanovic:2006xc}).
   We see that already at $M \approx 360\,\mbox{MeV}$ the
term $k_5 M^5 \ln(M/m_N)$ becomes as large as the leading
nonanalytic term at one-loop order, $k_2 M^3$, indicating the
importance of the fifth-order terms at unphysical pion masses.
   Since the contribution at order ${\cal O}(M^6)$ depends on a
number of unknown LECs, we do not attempt to perform a chiral
extrapolation up to this order here, but restrict the discussion
on the pion mass dependence of the term $k_7 M^6 \ln^2(M/m_N)$.
   Figure~\ref{Mass:k3k7Low} shows this dependence for pion masses below
$400\,\mbox{MeV}$.
   No errors are given for the LECs $c_1$, $c_2$, and $c_3$ in
Ref.~\cite{Becher:2001hv}.
   For an estimate we have assumed the relative errors of these
LECs and of $\texttt{g}_A$ to be $20\%$, and the corresponding
error for $k_7 M^6 \ln^2(M/m_N)$ is shown in
Fig.~\ref{Mass:k3k7Low}.
   For comparison we also show the nonanalytic term at fourth
order, $k_3 M^4 \ln(M/m_N)$.
   As expected, and in contrast to the fifth-order term, the two-loop term $k_7 M^6 \ln^2(M/m_N)$
is smaller than the one-loop contribution $k_3 M^4 \ln(M/m_N)$ in
the considered pion mass region.
   Note that the relative difference in the pion mass dependence
between $k_5 M^5 \ln(M/m_N)$ and $k_2 M^3$, as well as $k_7 M^6
\ln^2(M/m_N)$ and $k_3 M^4 \ln(M/m_N)$ is proportional to a factor
$M^2 \ln(M/m_N)$, and that for the physical pion mass the
differences in the two cases are comparable on an absolute scale.
\begin{figure}
\begin{center}
\epsfig{file=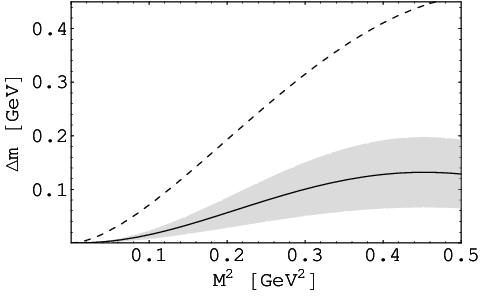,width=0.5\textwidth}
\end{center}
\caption{Pion mass dependence of the term $k_7 M^6 \ln^2(M/m_N)$
(solid line) for $M < 700\,\mbox{MeV}$. The shaded band
corresponds to relative errors of $20\%$ in the LECs. For
comparison also the term $k_3 M^4 \ln(M/m_N)$ (dashed line) is
shown.\label{Mass:k3k7High}}
\end{figure}
   We also show the pion mass dependence of the
terms $k_7 M^6 \ln^2(M/m_N)$ and $k_3 M^4 \ln(M/m_N)$ up to
$M\approx 700\, \mbox{MeV}$, which, however, is beyond the domain
that is considered suitable for the application of
Eq.~(\ref{Mass:Exp}).
   Again the sixth-order term remains much smaller than the
fourth-order one, also at higher pion masses.
   However, the above considerations are not reliable predictions
for the behavior of the complete two-loop contributions at
unphysical quark masses.
   This is because here only one of the terms at order ${\cal O}(q^6)$ is
considered, and the contribution of the analytic term proportional
to $M^6$ can be considerably larger than $k_7 M^6 \ln^2(M/m_N)$
depending on the values of the unknown LECs.

\subsection{Nucleon $\sigma$ term}

   The Feynman-Hellmann theorem \cite{Hellmann:1933,Feynman:1939} relates the nucleon mass to the
value of the nucleon scalar form factor at zero momentum transfer,
the so-called $\sigma$ term (see, e.g.,
\cite{Reya:1974gk,Pagels:1974se}),
\begin{equation}\label{Mass:FHTheo}
    \sigma(q^2=0) = M^2 \frac{\partial m_N}{\partial M^2}\,.
\end{equation}
   Applying the Feynman-Hellmann theorem to Eq.~(\ref{Mass:Exp}), the chiral expansion of $\sigma(0)$
is given by
\begin{align}\label{Mass:Sigma}
  \sigma(0) &= k_1 M^2 + \frac{3}{2}\, k_2 M^3 +2k_3 M^4 \logM
+\left(\frac{k_3}{2}+2k_4 \right)M^4 +\frac{5}{2}\,k_5 M^5 \logM  \notag \\
  & \quad +\frac{1}{2}\,(k_5+5k_6)M^5 + 3k_7 M^6 \logMsq + (k_7 +3k_8) M^6 \logM
+ \left( \frac{k_8}{2}+3k_9 \right)M^6.\notag\\
\end{align}
   The first four terms have already been determined in
Ref.~\cite{Becher:1999he}.
   To estimate the contributions of the terms of order ${\cal O}(M^5)$
we use the same values for the LECs as above, in particular the
value of $d_{16}$ as extracted from the reaction $\pi N
\rightarrow \pi \pi N$.
   The combined contributions at order ${\cal O}(M^5)$ are
\begin{equation}\label{Mass:M5Sigma}
   \frac{5}{2}\,k_5 M^5 \logM +\frac{1}{2}\,(k_5+5k_6)M^5 \approx
-0.2\,\mbox{MeV}.
\end{equation}
   Due to the dependence of the order ${\cal O}(M^6)$ nucleon mass contribution
on the specific values of the unknown LECs $e_i$, we do not
attempt to evaluate the terms at order ${\cal O}(M^6)$ in
Eq.~(\ref{Mass:Sigma}).

\section{Summary}

   We have shown details of how to consistently renormalize two-loop diagrams
in manifestly Lorentz-invariant BChPT within the framework of
infrared regularization.
   The renormalization procedure preserves all relevant symmetries
such that renormalized expressions fulfill the relevant Ward
identities.
   Renormalized diagrams also obey the standard power counting.
   We have presented a simplified method of renormalizing
diagrams with one small scale, which relies on dimensional
analysis.
   Integrals of this kind appear, e.g., in the calculation of the
nucleon mass or the axial-vector coupling constant.
   In this method integrals derived from the original
expressions are renormalized using the $\widetilde{{\rm MS}}$
scheme, which simplifies the calculations considerably.
   As an application we have calculated the nucleon mass up to and
including order ${\cal O}(q^6)$.
   For physical values of the pion mass, the numerical estimate of
the two-loop contributions is reasonably small.
   For example, the estimate of the ${\cal O}(q^5)$ term of the
nucleon $\sigma$ term is $-0.2\,\mbox{MeV}$.
   However, when considering the nucleon mass as a function of the pion mass, we
have seen that already at a pion mass of $360\,\mbox{MeV}$ the
nonanalytic contribution at order ${\cal O}(q^5)$ may become as
large as the nonanalytic ${\cal O}(q^3)$ contribution.
   From this one cannot conclude that the chiral expansion breaks down at
this value of the pion mass since the analytic terms at ${\cal
O}(q^5)$ might cancel parts of the nonanalytic term.
   One should, however, take special care when performing chiral
extrapolations beyond this value.

\begin{acknowledgments}
M.~R.~Schindler, D.~Djukanovic, and J.~Gegelia  acknowledge
support by the Deutsche Forschungsgemeinschaft (SFB 443 and SCHE
459/2-1). This work was partially supported by the US Department of Energy under grant DE-FG02-93ER40756.

\end{acknowledgments}

\appendix

\section{Dimensional counting analysis}\label{App:DimCount}
   Analytic expressions for two-loop integrals, especially when two mass
scales such as the pion mass $M$ and the nucleon mass in the
chiral limit $m$ appear in the same integral, can be extremely
difficult to obtain.
   Since we are interested in the chiral expansion of
the considered integrals in the present work, we do not have to
find a closed-form solution to the appearing integrals, but can
use a method called dimensional counting analysis
\cite{Gegelia:1994zz} for the evaluation of integrals.
   A closely related way of calculating loop integrals is the so-called
"strategy of regions" \cite{Smirnov:2002pj}.
   Here we present an illustration of dimensional counting for one- and
two-loop integrals.

\subsection{One-loop integrals}
   The advantage of the dimensional counting analysis for one-loop
integrals lies in its applicability to dimensionally regulated
integrals containing several different masses.
   Consider integrals with two different mass scales, $M$ and $m$,
where $M<m$, and a possible external momentum $p$ with $p^2\approx
m^2$.
   Dimensional counting provides a method to reproduce the
expansion of the integral for small values of $M$ at fixed $p^2-
m^2$.
   To that end one rescales the loop momentum $k\mapsto M^{\alpha_i}
\tilde{k}$, where $\alpha_i$ is a non-negative real number.
   After extracting an overall factor of $M$ one expands the
integrand in positive powers of $M$ and interchanges summation and
integration.
   The sum of all possible rescalings with subsequent expansions with
nontrivial coefficients then reproduces the expansion of the
result of the original integral.

   To be specific, consider the integral
\begin{equation}\label{DimC:Int}
H_{\pi N}(p^2)=
\frac{i}{(2\pi)^n}\int\frac{d^nk}{(k^2-M^2+i0^+)[(k+p)^2-m^2+i0^+]}.
\end{equation}
   It can be evaluated analytically and the result is given in App.~\ref{App:IntDef}.
   After rescaling one obtains
\begin{equation}\label{DimC:IntRescaled}
H_{\pi N}(p^2) \mapsto \frac{i}{(2\pi)^n}\int\frac{M^{n \alpha_i}
d^n\tilde{k}}{[\tilde{k}^2 M^{2 \alpha_i}-M^2+i0^+][\tilde{k}^2
M^{2 \alpha_i}+2 p\cdot \tilde{k} M^{\alpha_i}+ p^2-m^2+i0^+]}.
\end{equation}
   No overall factor of $M$ can be extracted from the second
propagator, which is therefore expanded in positive powers of $M$.
   As a result the integration variable $\tilde{k}$ only appears
in positive powers in the expanded expression of this propagator.
   If $0<\alpha_i<1$ one can extract the factor $M^{-2\alpha_i}$ from the
first propagator, which takes the form
\begin{equation}\label{DimC:aless1}
    \frac{1}{\tilde{k}^2-M^{2-2 \alpha_i}+i 0^+}\,.
\end{equation}
   Expanding in positive powers of $M$ and interchanging summation
and integration one obtains integrals of the type
\begin{equation}\label{DimC:aless1exp}
    \int d^n\tilde{k} \frac{1}{(\tilde{k}^2+i0^+)^j}\,.
\end{equation}
   Combined with the expansion of the second propagator the
resulting coefficients in the expansion in $M$ are integrals of
the type
\begin{equation}\label{DimC:aless1exp2}
    \int d^n\tilde{k} \frac{\tilde{k}^m}{(\tilde{k}^2+i0^+)^j}\,,
\end{equation}
which vanish in dimensional regularization.
   For the case $1<\alpha_i$ the first propagator in
Eq.~(\ref{DimC:IntRescaled}) can be rewritten as
\begin{equation}\label{DimC:agreater1}
\frac{1}{M^{2}} \ \frac{1}{(\tilde k^2 M^{2 \alpha_i-2}-1+i
0^+)}\,.
\end{equation}
   Expanding in $M$ and combining with the expansion of the second
propagator one obtains integrals of the type
\begin{equation}\label{DimC:agreater1exp}
    \int d^n\tilde{k}\, \tilde{k}^j\,,
\end{equation}
which, again, vanish in dimensional regularization.
   The only contributions to $H_{\pi N}(p^2)$ can therefore stem
from $\alpha_i=0$ and $\alpha_i=1$.
   For $\alpha_i=0$ one obtains
\begin{equation}\label{DimC:aeq0}
H_{\pi N}^{(0)}(p^2)=\frac{i}{(2\pi)^n}\sum_{i=0}^{\infty} \left(
M^2\right)^i \int \frac{d^nk}{[k^2
+i0^+]^{1+i}[(k+p)^2-m^2+i0^+]}\,,
\end{equation}
   while the expression for $\alpha_i=1$ reads
\begin{equation}\label{DimC:aeq1}
H_{\pi N}^{(1)}(p^2)=\frac{i}{(2\pi)^n}\sum_{i=0}^{\infty}
(-1)^i\frac{M^{n-2+i}}{(p^2-m^2)^{1+i}} \int\frac{d^n\tilde{k}
(\tilde{k}^2 M+2 p\cdot\tilde{k})^i}{[\tilde{k}^2-1+i0^+]}.
\end{equation}
   The expansion of $H_{\pi N}(p^2)$ is then given by
\begin{equation}\label{DimC:Sum}
    H_{\pi N}(p^2)=H_{\pi N}^{(0)}(p^2)+H_{\pi N}^{(1)}(p^2),
\end{equation}
which correctly reproduces the result of App.~\ref{App:IntDef}.

\subsection{Two-loop integrals}\label{App:DimCountTL}
   While one of the advantages of the dimensional counting
method lies in its applicability to integrals containing several
mass scales, a difficulty arises for the calculation of the
nucleon mass.
   Since integrals have to be evaluated on-mass-shell, the two
small scales $M$ and $p^2-m^2$ are not independent of each other
and are comparable in size.
   Therefore an expansion in $\frac{M}{p^2-m^2}$ does not
converge.
   By the choice of the nucleon propagator mass to include all
contact interaction contributions, the terms $p^2-m^2$ in the
propagator can be neglected in two-loop integrals since they are
of higher order in the loop expansion.
   The two-loop integrals contributing to the nucleon mass are
therefore reduced to integrals with only one small mass scale, for
which an expansion in $M$ can be obtained.

   For the extension of the dimensional counting method to two-loop
integrals
\begin{eqnarray}\label{DimC:TLDef}
   \lefteqn{ H_2(a,b,c,d,e|n) =  \iint\frac{d^nk_1d^nk_2}{(2\pi)^{2n}}
\frac{1}{[k_1^2-M^2+i0^+]^a[k_2^2-M^2+i0^+]^b} } \nonumber\\
   &&\times \frac{1}{[k_1^2+2p\cdot
k_1+i0^+]^c [k_2^2+2p\cdot k_2+i0^+]^d[(k_1+k_2)^2+2p\cdot
k_1+2p\cdot
k_2+i0^+]^e}\,,\nonumber\\
\end{eqnarray}
 one has to consider all possible
combinations of rescaling the integration variables $k_1\mapsto
M^{\alpha_i}\tilde{k}_1$, $k_2\mapsto M^{\beta_i}\tilde{k}_2$.
   The expansion of the two-loop integral is then given by
\begin{align}\label{DimC:TLResult}
    H_2(a,b,c,d,e|n) &=\sum_{\alpha_i,\beta_i}M^{\varphi(\alpha_i,\beta_i)}h^{(\alpha_i,\beta_i)}(a,b,c,d,e|n) \notag\\
    &= \sum_{\alpha_i,\beta_i}H^{(\alpha_i,\beta_i)}(a,b,c,d,e|n),
\end{align}
where $\varphi(\alpha_i,\beta_i)$ is the overall power of $M$
extracted for each rescaling, the functions
$h^{(\alpha_i,\beta_i)}(a,b,c,d,e|n)$ are the expressions for the
integrated expansions, and we have defined
$H^{(\alpha_i,\beta_i)}(a,b,c,d,e|n)=M^{\varphi(\alpha_i,\beta_i)}h^{(\alpha_i,\beta_i)}(a,b,c,d,e|n)$
to simplify the notation.
   Following the discussion of the one-loop sector one sees that
the only combinations $(\alpha_i,\beta_i)$ that give non-vanishing
contributions are $(0,0)$, $(1,0)$, $(0,1)$ and $(1,1)$, so that a two-loop integral is given by
\begin{align}\label{DimC:F1F2F3F4}
   H_2(a,b,c,d,e|n) &= H^{(0,0)}(a,b,c,d,e|n)+H^{(1,0)}(a,b,c,d,e|n)+H^{(0,1)}(a,b,c,d,e|n)\notag\\
&\quad +H^{(1,1)}(a,b,c,d,e|n).
\end{align}
To shorten the notation and to avoid confusion with other superscripts used for further expansions in this paper the contributions corresponding to $H^{(0,0)}$, $H^{(1,0)}$, $H^{(0,1)}$, and $H^{(1,1)}$ are also denoted by $F_1$, $F_2$, $F_3$ and $F_4$, respectively.

   From a technical point of view it is convenient to consider the
rescaling $k_1\mapsto (M/m)^{\alpha_i}\tilde{k}_1$, $k_2\mapsto
(M/m)^{\beta_i}\tilde{k}_2$, since then the integration variables
$\tilde{k}$ have dimension of momenta.
   This also facilitates the evaluation of certain loop integrals
appearing in the calculation of the nucleon mass.

   As an example consider the integral $H_2(1,1,1,1,1|n)$.
   For $(0,0)$ the resulting integrals read
\begin{align}\label{DimC:TL00}
   H^{(0,0)}(1,1,1,1,1|n) &=\sum_{i,j}M^{2i+2j}\iint\frac{d^nk_1d^nk_2}{(2\pi)^{2n}}
\frac{1}{[k_1^2+i0^+]^{1+i}[k_2^2+i0^+]^{1+j}[k_1^2+2p\cdot k_1+i0^+]  } \notag\\
   &\times\frac{1}{
[k_2^2+2p\cdot k_2+i0^+][(k_1+k_2)^2+2p\cdot k_1+2p\cdot
k_2+i0^+]}\,.\nonumber\\
\end{align}
   While still a two-loop integral that does not directly factorize into
the product of one-loop integrals, the vanishing of the mass scale
$M$ simplifies the evaluation of the integral.
   The rescaling of only $k_1$ leads to
\begin{align}\label{DimC:TL10}
   H^{(1,0)}(1,1,1,1,1|n) &= \sum_{i,j,l}(-1)^{j+l} M^{n-3+2i+j+l}m^{3-n-j-l} \iint
\frac{d^n\tilde{k}_1d^nk_2}{(2\pi)^{2n}} \frac{1}{[\tilde{k}_1^2-m^2+i0^+]}  \notag\\
   & \times
   \frac{(\tilde{k}_1^2)^j(\frac{M}{m} \tilde{k}_1^2+2p\cdot \tilde{k}_1+2\tilde{k}_1\cdot k_2)^l}{
[k_2^2+i0^+]^{1+i}[2p\cdot \tilde{k}_1+i0^+]^{1+j}[k_2^2+2p\cdot
k_2+i0^+]^{2+l}}\,,\nonumber\\
\end{align}
while the expression for $H^{(0,1)}(1,1,1,1,1|n)$ can be obtained by substituting
$\tilde{k}_1\mapsto \tilde{k}_2$ and $k_2\mapsto k_1$ in
Eq.~(\ref{DimC:TL10}).
   One sees that the integrals of Eq.~(\ref{DimC:TL10})
can be reduced to the product of tensorial one-loop integrals,
which is a considerable simplification compared to the original
integral.
   The last contribution stems from $\alpha_i=1$, $\beta_i=1$ and
reads
\begin{align}\label{DimC:TL11}
   H^{(1,1)}(1,1,1,1,1|n) &= \sum_{i,j,l}(-1)^{i+j+l} \left(\frac{M}{m}\right)^{2n-7+i+j+l}
\iint
\frac{d^n\tilde{k}_1d^n\tilde{k}_2}{(2\pi)^{2n}}\frac{1}{[\tilde{k}_1^2-m^2+i0^+]}  \notag\\
    &\hspace{-3em} \times \frac{(\tilde{k}_1^2)^i(\tilde{k}_2^2)^j(\tilde{k}_1^2+2\tilde{k}_1\cdot\tilde{k}_2+\tilde{k}_2^2)^l}
{[\tilde{k}_2^2-m^2+i0^+][2p\cdot \tilde{k}_1+i0^+]^{1+i}[2p\cdot
\tilde{k}_2+i0^+]^{1+j}[2p\cdot \tilde{k}_1+2p\cdot
\tilde{k}_2+i0^+]^{1+l}}\,,\nonumber\\
\end{align}
   where the integration can be reduced to the evaluation of a set
of basis integrals (see App.~\ref{App:IntEv}).
   The sum of all four contributions reproduces the $M$ expansion
of the integral $H_2(1,1,1,1,1|n)$,
\begin{align}\label{DimC:TLtotal}    H_2(1,1,1,1,1|n) & = H^{(0,0)}(1,1,1,1,1|n)+H^{(1,0)}(1,1,1,1,1|n)+ H^{(0,1)}(1,1,1,1,1|n) \notag\\
&\quad +H^{(1,1)}(1,1,1,1,1|n).
\end{align}

\section{Integrals at the one-loop level}\label{App:IntDef}

   Using dimensional regularization \cite{'tHooft:1972fi} the one-loop integrals
are defined as
\begin{align*}
H_{\pi} &= i\int\frac{d^nk}{(2\pi)^n} \frac{1}{k^2-M^2+i0^+},\\
g^{\mu\nu}H_\pi^{(00)} &= i\int\frac{d^nk}{(2\pi)^n} \frac{k^\mu k^\nu}{k^2-M^2+i0^+},\\
\left(g^{\mu\nu}g^{\rho\sigma}\!+g^{\mu\rho}g^{\nu\sigma}\!+g^{\mu\sigma}g^{\nu\rho}\right)
   \!H_\pi^{(0000)} &= i\int\frac{d^nk}{(2\pi)^n} \frac{k^\mu k^\nu k^\rho k^\sigma}{k^2-M^2+i0^+},\\
H_{\pi\pi} &= i\int\frac{d^nk}{(2\pi)^n} \frac{1}{[k^2-M^2+i0^+]^2},\\
g^{\mu\nu}H_{\pi\pi}^{(00)} &= i\int\frac{d^nk}{(2\pi)^n} \frac{k^\mu k^\nu}{[k^2-M^2+i0^+]^2},\\
H_N &= i\int\frac{d^nk}{(2\pi)^n} \frac{1}{k^2-m^2+i0^+},\\
H_{{\pi}N}(p^2)&=
   i\int\frac{d^nk}{(2\pi)^n}
   \frac{1}{[k^2-M^2+i0^+][(k+p)^2-m^2+i0^+]},\\
p^{\mu}H_{{\pi}N}^{(p)}(p^2) &=
   i\int\frac{d^nk}{(2\pi)^n}
   \frac{k^{\mu}}{[k^2-M^2+i0^+][(k+p)^2-m^2+i0^+]},\\
H_{{\pi\pi}N}(p^2) &= i\int\frac{d^nk}{(2\pi)^n} \frac{1}
   {[k^2-M^2+i0^+]^2[(k+p)^2-m^2+i0^+]},\\
p^{\mu}H_{{\pi\pi}N}^{(p)}(p^2) &= i\int\frac{d^nk}{(2\pi)^n}
\frac{k^\mu}
   {[k^2-M^2+i0^+]^2[(k+p)^2-m^2+i0^+]}.
\end{align*}
   The tensorial loop integrals can be reduced to scalar ones \cite{Passarino:1978jh} and
we obtain
\begin{align*}
H_\pi^{(00)} &= \frac{M^2}{n} H_\pi,\\
H_{\pi\pi}^{(00)} &= \frac{1}{n}\left[H_\pi + M^2 H_{\pi\pi}
\right],\\
H_\pi^{(0000)} &= \frac{M^4}{n(n+2)}\,H_\pi,\\
H_{{\pi}N}^{(p)}(p^2) &= \frac{1}{2p^2}
   \left[H_{\pi}-H_N-(p^2-m^2+M^2)H_{{\pi}N}(p^2)\right],\\
H_{{\pi\pi}N}^{(p)}(p^2) &= \frac{1}{2p^2}
   \left[H_{\pi\pi}-H_{\pi N}-(p^2-m^2+M^2)H_{{\pi\pi}N}(p^2)\right].
\end{align*}

Defining
\begin{displaymath}
\lambda =\frac{\mu^{n-4}}{16\pi^2}\left\{
\frac{1}{n-4}-\frac{1}{2} \left[ \ln (4\pi)
-\gamma_E+1\right]\right\},
\end{displaymath}
where $\gamma_E=-\Gamma '(1)$ is Euler's constant, and
\begin{displaymath}
\Omega=\frac{p^2-m^2-M^2}{2mM},
\end{displaymath}
the scalar loop integrals are given by \cite{Fuchs:2003qc}
\begin{align*}
   H_\pi &=
2M^2\lambda+\frac{M^2}{8\pi^2}\logM,\\
   H_{\pi\pi} &= 2\lambda+\frac{1}{16\pi^2}\left[1+2\logM\right] ,\\
   H_N &= 2m^2\lambda +\frac{m^2}{8\pi^2}\ln \frac{m}{\mu},\\
   H_{\pi N}(p^2) &= 2\lambda+\frac{1}{16\pi^2}\left[-1
+\frac{p^2-m^2+M^2}{p^2}\logM +\frac{2mM}{p^2}F(\Omega)\right],
\end{align*}
where
\begin{align*}
F(\Omega) &= \left \{ \begin{array}{ll}
\sqrt{\Omega^2-1}\ln\left(-\Omega-\sqrt{\Omega^2-1}\right),&\Omega\leq -1,\\
\sqrt{1-\Omega^2}\arccos(-\Omega),&-1\leq\Omega\leq 1,\\
\sqrt{\Omega^2-1}\ln\left(\Omega+\sqrt{\Omega^2-1}\right)
-i\pi\sqrt{\Omega^2-1},&1\leq \Omega.
\end{array} \right.
\end{align*}
   The integral $H_{\pi\pi N}$ can be obtained from $H_{\pi
N}(p^2)$ by differentiating with respect to $M^2$.

\section{Integrals at the two-loop level}\label{App:TwoLoopIntDef}

   We define the one-loop integral
\begin{displaymath}
   H_{{\pi}N}(a,b|n)=\int\frac{d^nk}{(2\pi)^n}\frac{1}{[k^2-M^2+i0^+]^a[k^2+2p\cdot
   k+i0^+]^b}.
\end{displaymath}
   Note that we have not included a factor $i$ in the definition,
since the one-loop integrals $H_{{\pi}N}(a,b|n)$ always appear in
the product $H_{{\pi}N}(a_1,b_1|n_1)H_{{\pi}N}(a_2,b_2|n_2)$.

   The two-loop integrals $H_2(a,b,c,d,e|n)$ are defined as
\begin{displaymath}
    H_2(a,b,c,d,e|n)=\iint \frac{d^nk_1 d^nk_2}{(2\pi)^{2n}} \, \frac{1}{A^a B^b C^c D^d
    E^e}\,,
\end{displaymath}
where
\begin{eqnarray*}
  A &=& k_1^2-M^2+i0^+, \nonumber\\
  B &=& k_2^2-M^2+i0^+, \nonumber\\
  C &=& k_1^2+2p\cdot k_1+i0^+, \nonumber\\
  D &=& k_2^2+2p\cdot k_2+i0^+, \nonumber\\
  E &=& k_1^2+2p\cdot k_1+2k_1\cdot k_2+2p\cdot k_2+k_2^2+i0^+.
\end{eqnarray*}
   The product of two one-loop integrals with the same space-time dimension
$n$ can then be written as
\begin{displaymath}
   H_{{\pi}N}(a_1,b_1|n)H_{{\pi}N}(a_2,b_2|n)=H_2(a_1,a_2,b_1,b_2,0|n).
\end{displaymath}
   We do not attempt to evaluate the integrals $H_2(a,b,c,d,e|n)$
here.
   Instead the expressions for the integrals relevant to
the nucleon mass including the corresponding counterterm
contributions are given in App.~\ref{App:IntEv}.

   Tensorial integrals have been reduced to scalar ones in the
same dimension using methods similar to the one-loop integrals, or
the following relations to scalar integrals in higher dimensions
have been used:\allowdisplaybreaks[1]
\begin{align*}
&H_2^{\mu,}(a,b,c,d,e|n) = \iint \frac{d^nk_1
d^nk_2}{(2\pi)^{2n}} \frac{k_1^{\mu}} {A^a B^b C^c D^d E^e}  \\
   &\quad= -16\pi^2 p^\mu\, \Big[b\,c\,H_2\left(a,b+1,c+1,d,e|n+2\right)
+c\,d\,H_2\left(a,b,c+1,d+1,e|n+2\right) \\
   &\qquad+c\,e\,H_2\left(a,b,c+1,d,e+1|n+2\right)
+b\,e\,H_2\left(a,b+1,c,d,e+1|n+2\right)\Big],\\
  &H_2^{\mu\nu ,}(a,b,c,d,e|n) = \iint \frac{d^nk_1 d^nk_2}{(2\pi)^{2n}}
\frac{k_1^{\mu}k_1^{\nu}} {A^a B^b C^c D^d E^e}  \\
  &\quad= \frac{(4\pi)^2}{2}\,g^{\mu\nu}
\left[b\, H_2(a,b+1,c,d,e|n+2)
+ d\, H_2(a,b,c,d+1,e|n+2)\right.\nonumber\\
  &\qquad\left. +
e \, H_2(a,b,c,d,e+1|n+2)\right]+{\cal O}(p),
  \\
  & H_2^{\mu\nu\lambda ,}(a,b,c,d,e|n) = \iint \frac{d^nk_1 d^nk_2}{(2\pi)^{2n}}
\frac{k_1^{\mu}k_1^{\nu}k_1^{\lambda}} {A^a B^b C^c D^d E^e}  \\
  &\quad= -\frac{(4\pi)^4}{2}\,
\left[g^{\mu\nu}p^{\lambda}+g^{\mu\lambda}p^{\nu}+g^{\nu\lambda}p^{\mu}\right]
\left[b\,(b+1)\, c \, H_2(a,b+2,c+1,d,e|n+4)\right. \\
  &\qquad+\left.b\,(b+1)\,e \, H_2(a,b+2,c,d,e+1|n+4)\right]
+{\cal O}(p^3),
  \\
  & H_2^{\mu ,\nu}(a,b,c,d,e|n) = \iint \frac{d^nk_1
d^nk_2}{(2\pi)^{2n}}
\frac{k_1^{\mu}k_2^{\nu}} {A^a B^b C^c D^d E^e} \\
  &\quad= -\frac{(4\pi)^2}{2}\,
g^{\mu\nu}\, e\, H_2(a,b,c,d,e+1|n+2)+\mathcal{O}(p),
  \\
  & H_2^{\mu ,\alpha\beta}(a,b,c,d,e|n) = \iint \frac{d^nk_1
d^nk_2}{(2\pi)^{2n}}
\frac{k_1^{\mu}k_2^{\alpha}k_2^{\beta}} {A^a B^b C^c D^d E^e}  \\
  &\quad= -\frac{(4\pi)^2}{2}\,g^{\alpha\beta}p^{\mu}
\left[c\,H_2(a,b,c+1,d,e|n+2)
+e\,H_2(a,b,c,d,e+1|n+2)\right]  \\
  &\qquad  + \frac{(4\pi)^4}{2}\,
\left[g^{\alpha\beta}p^{\mu}+g^{\mu\alpha}p^{\beta}+g^{\mu\beta}p^{\alpha}\right]
\left[a\,e\,(e+1)\,
H_2(a+1,b,c,d,e+2|n+4)\right.\nonumber\\
  &\qquad  +a\,d\,e\,
H_2(a+1,b,c,d+1,e+1|n+4)\nonumber\\
  &\qquad  +c\,d\,e\,
H_2(a,b,c+1,d+1,e+1|n+4)\nonumber\\
  &\qquad\left.
+d\,e\,(e+1)\, H_2(a,b,c,d+1,e+2|n+4)\right]+{\cal O}(p^3),
  \\
  & H_2^{\alpha\beta , \mu\nu}(a,b,c,d,e|n) = \iint
\frac{d^nk_1 d^nk_2}{(2\pi)^{2n}}
\frac{k_1^{\alpha}k_1^{\beta}k_2^{\mu}k_2^{\nu}} {A^a B^b C^c D^d
E^e}  \\
  &\quad= \frac{(4\pi)^4}{4}\,
\left[g^{\alpha\beta}g^{\mu\nu}+g^{\alpha\mu}g^{\beta\nu}+g^{\alpha\nu}g^{\beta\mu}\right]
e\,(e+1)\,H_2(a,b,c,d,e+2|n+4)\nonumber\\
&\qquad + \frac{(4\pi)^2}{4}\,
g^{\alpha\beta}g^{\mu\nu}H_2(a,b,c,d,e|n+2)+\mathcal{O}(p).
\end{align*}
   Here, ${\cal O}(p)$ stands for terms proportional to $p^{\, \rho}$,
where $\rho$ denotes the Lorentz index corresponding to the
integral under consideration.
   In our calculation of the nucleon mass these terms appear in
combination with expressions like $(\psl-m)\gamma_\rho(\psl+m)$,
resulting in higher-order contributions that are not considered.

\section{Evaluation of two-loop integrals}\label{App:IntEv}

   As seen in Sec.~\ref{Sec:TwoLoop} the calculation of the two-loop
integrals relevant to the nucleon mass reduces to the evaluation
of the $F_4$ part of the respective integrals.
   The $F_4$ parts are sums of tensor integrals, which can be
reduced to scalar integrals \cite{Passarino:1978jh} of the form
\begin{eqnarray}\label{IntEv:GenInt}
   H_2^{(1,1)}(a,b,c,d,e|n)\hspace{-0.5em}&=&\hspace{-0.5em}\iint \frac{d^nk_1d^nk_2}{(2\pi)^{2n}} \frac{1}
{[k_1^2-m^2+i0^+]^a[k_2^2-m^2+i0^+]^b[2p\cdot k_1+i0^+]^c}\nonumber\\
   && \times \frac{1}{[2p\cdot
k_2+i0^+]^d[2p\cdot k_1+2p\cdot k_2+i0^+]^e}\,,
\end{eqnarray}
   where the superscript $(1,1)$ indicates that these integrals
have been obtained after rescaling both integration variables (see
App.~\ref{App:DimCount}) and $a,b,c,d,e$ are integers.
   Depending on the values of the exponents $c$, $d$, and $e$, one
can evaluate $H_2^{(1,1)}(a,b,c,d,e|n)$ with the help of several
basic integrals.

\subsection{$e=0$}
   If the exponent $e$ vanishes, the integral can be written as
the product of one-loop integrals,
\begin{equation}\label{IntEv:e0}
    H_2^{(1,1)}(a,b,c,d,0|n)=H_1^{(1)}(a,c|n)H_1^{(1)}(b,d|n),
\end{equation}
where
\begin{eqnarray}\label{IntEv:HpiN}
    H_1^{(1)}(a,b|n)&=&\int \frac{d^nk}{(2\pi)^n} \frac{1}{[k^2-m^2+i0^+]^a[2p\cdot
    k+i0^+]^b} \nonumber \\
    &=& \frac{i^{1-2a-2b}}{2^b (4\pi)^{n/2}}
\frac{\Gamma[\frac{1}{2}]\Gamma[a+\frac{b}{2}-\frac{n}{2}]}{\Gamma[a]\Gamma[\frac{b+1}{2}]}
\, (m^2)^{n/2-a-b/2}(p^2)^{-b/2}\,.
\end{eqnarray}

\subsection{$c=d=0, e\neq0$}
   If $c=d=0$, the expression for the integral $H_2(a,b,0,0,e|n)$ reads
\begin{eqnarray}\label{IntEv:cd0}
   \lefteqn{ H_2^{(1,1)}(a,b,0,0,e|n) } \nonumber\\
   &=& \hspace{-0.5em} \iint \frac{d^nk_1d^nk_2}{(2\pi)^{2n}} \frac{1}
{[k_1^2-m^2+i0^+]^a[k_2^2-m^2+i0^+]^b[2p\cdot k_1+2p\cdot k_2+i0^+]^e}  \nonumber\\
  &=& \hspace{-0.5em} \frac{i^{2-2a-2b-2e}}{2(4\pi)^n}\,(m^2)^{n-a-b-e/2}(p^2)^{-e/2}
  \frac{\Gamma[a+\frac{e}{2}-\frac{n}{2}]\Gamma[b+\frac{e}{2}-\frac{n}{2}]
\Gamma[a+b+\frac{e}{2}-n]\Gamma[\frac{e}{2}]}
{\Gamma[a]\Gamma[b]\Gamma[e]\Gamma[a+b+e-n]}.\nonumber\\
\end{eqnarray}

\subsection{$d=0,\; c\neq0,\;e\neq0$ and $c=0,\; d\neq0,\; e\neq0$}
   For vanishing $d$ with non-vanishing $c$ and $e$ we consider the integral
$H_2^{(1,1)}(a,b,c,0,e|n)$ for $p^2=m^2$,
\begin{eqnarray}\label{IntEv:d0massshell}
   H_2^{(1,1)}(a,b,c,d,e|n) \hspace{-.5em}&=& \hspace{-.5em}\iint \frac{d^nk_1d^nk_2}{(2\pi)^{2n}} \frac{1}
{[k_1^2-m^2+i0^+]^a[k_2^2-m^2+i0^+]^b[2p\cdot k_1+i0^+]^c}\nonumber\\
   && \times \left.\frac{1}{[2p\cdot
k_2+i0^+]^d[2p\cdot k_1+2p\cdot k_2+i0^+]^e}\right|_{p^2=m_N^2}\,.
\end{eqnarray}
   Note that the mass terms $m$ in the first two propagators stem
from the rescaling of the loop momenta, while we have to consider
$p^2=m_N^2$ when evaluating the nucleon mass.
   In the calculations performed in this work the difference between
$p^2=m^2$ and $p^2=m_N^2$ in these integrals is of higher order.

   The result for $H_2^{(1,1)}(a,b,c,0,e|n)$ is given by the sum
\begin{equation}\label{IntEv:d0}
    H_2^{(1,1)}(a,b,c,0,e|n)=\sum_{l=0}^{c-1}{c-1 \choose l}(-1)^l
    Z_{(e+l-2)/2}(a,b,c,0,e|n),
\end{equation}
where
\begin{eqnarray}\label{IntEv:Z}
   \lefteqn{\hspace{-2em}Z_{\alpha}(a,b,c,0,e|n)=
   \frac{i^{2-2a-2b-2c-2e}}{4(4\pi)^n}\, m^{2n-2a-2b-2c-2e} \frac{\Gamma[\alpha+1]}{\Gamma[\alpha+2]}\,
}\nonumber\\
   && \times \frac{\Gamma[a+\frac{c}{2}+\frac{e}{2}-\frac{n}{2}]\Gamma[b+\frac{c}{2}+\frac{e}{2}-\frac{n}{2}]
\Gamma[\frac{c}{2}+\frac{e}{2}]\Gamma[a+b+\frac{c}{2}+\frac{e}{2}-n]}
{\Gamma[a]\Gamma[b]\Gamma[c]\Gamma[e]\Gamma[a+b+c+e-n]}
\nonumber\\
   && \times {}_3F_2\left(\left.{1,c/2+e/2,a+c/2+e/2-n/2 \atop \alpha+2,a+b+c+e-n}\right|1\right)
\end{eqnarray}
and ${}_3F_2\left(\left.{a,b,c \atop d,e}\right|z\right)$ is a
hypergeometric function.
   The case $c=0$, $d\neq0$, $e\neq0$ is obtained by replacing $c$ with
   $d$ and interchanging $a$ and $b$ in Eq.~(\ref{IntEv:Z}).

\subsection{$c\neq0,\;d\neq0,\;e\neq0$}
   For the case that none of the exponents $c$, $d$, $e$ vanishes, it is
convenient to perform an expansion into partial fractions,
\begin{eqnarray}\label{IntEv:PartialFrac}
    \lefteqn{\frac{1}{[2p\cdot
k_2+i0^+][2p\cdot k_1+2p\cdot
k_2+i0^+]}}\nonumber\\
&&=\frac{1}{[2p\cdot k_1+i0^+][2p\cdot
k_2+i0^+]}-\frac{1}{[2p\cdot k_1+i0^+][2p\cdot k_1+2p\cdot
k_2+i0^+]}\,,\nonumber\\
\end{eqnarray}
until one obtains a sum of integrals of the form
$H_2^{(1,1)}(a,b,\tilde{c},0,\tilde{e})$ and
$H_2^{(1,1)}(a,b,\tilde{c},\tilde{d},0)$, which are evaluated as
described above.

\subsection{Subtraction terms}

   In addition to the integrals given above the evaluation of the
subintegrals for the $F_4$ terms requires the integrals
\begin{equation}\label{IntEv:piNomega}
H_1^{(1,1)}(a,b;\omega|n)=\int \frac{d^nk}{(2\pi)^n}
\frac{1}{[k^2-m^2+i0^+]^a[2p\cdot k+\omega+i0^+]^b}\,,
\end{equation}
   where $\omega=2p\cdot q$ with $q$ the second loop
momentum.
   The integral $H_1^{1,1}(a,b;\omega|n)$ is given by the sum
\begin{eqnarray}\label{IntEv:piNomegaResult}
   \lefteqn{H_1^{(1,1)}(a,b;\omega|n)} \nonumber\\
   &&= \frac{i^{1-2a-2b}m^{n-2a-2b}}{(4\pi)^{n/2}} \sum_{l=0}^{\infty}
\frac{\Gamma[\frac{b}{2}+\frac{l}{2}]\Gamma[a+\frac{b}{2}-\frac{n}{2}+\frac{l}{2}]}{2\Gamma[a]\Gamma[b]\Gamma[l+1]}
\, \left(\frac{\omega}{m^2}\right)^l \left(\frac{m^2}{p^2}\right)^{b/2+l/2}. \nonumber\\
\end{eqnarray}
   The sum contains an infinite number of terms.
   However, when performing the second loop integration over $q$ in the considered
counterterm integrals, increasing orders of $\omega=2p\cdot q$
contribute to increasing chiral orders.
   Therefore only a finite number of terms in Eq.~(\ref{IntEv:piNomegaResult}) is
needed in the calculation of the nucleon mass.

\subsection{Results for $H_2^{(1,1)}(a,b,c,d,e|n)$ and counterterm
integrals}\label{App:Subsec:Results}

   The results for the $H_2^{(1,1)}$ parts of the two-loop integrals
contributing to the nucleon mass evaluated on-mass-shell are given
by\allowdisplaybreaks
\begin{align*}
     &\tilde\mu^{8-2n}H_2^{(1,1)}(1,1,0,0,1|n) \\
  &\quad = -\frac{1}{\epsilon^2}\,\frac{3M^4}{1024\pi^4 m^2}
-\frac{1}{\epsilon}\,\frac{M^4}{1024\pi^4m^2}\left[ 1+12\logM
\right]-\frac{M^4}{2048\pi^4m^2}\left[ \pi^2+10  \right.\\
 &\left.\qquad+8\logM+48\logMsq
\right],\\
     &\tilde\mu^{8-2n}H_2^{(1,1)}(1,1,0,0,2|n+2) \\
 &\quad = \frac{1}{\epsilon^2}\,\frac{M^6}{24576\pi^6 m^2}
-\frac{1}{\epsilon}\,\frac{M^6}{36864\pi^6m^2}\left[ 1-6\logM \right] +\frac{M^6}{442368\pi^6m^2}\,\bigg[ 3\pi^2 \\
 &\qquad \left.+26-48\logM+144\logMsq \right],\\
     &\tilde\mu^{8-2n}H_2^{(1,1)}(1,2,0,0,2|n+2) \\
 &\quad = \frac{1}{\epsilon^2}\,\frac{M^4}{16384\pi^6m^2}
+\frac{1}{\epsilon}\,\frac{M^4}{4096\pi^6m^2}\logM
+\frac{M^4}{98304\pi^6m^2}\,\left[ \pi^2+6+48\logMsq \right] ,\\
     &\tilde\mu^{8-2n}H_2^{(1,1)}(1,1,1,0,2|n+2) \\
 &\quad = -\frac{1}{\epsilon^2}\,\frac{M^6}{98304\pi^6m^4}+\frac{1}{\epsilon}\,\left\{\frac{M^5}{12288\pi^5m^3}
+\frac{M^6}{147456\pi^6m^4}\,\left[ 1-6\logM \right] \right\}\\
 &\qquad+\frac{M^5}{36864\pi^5m^3}\,\left[6\ln(2)-5+12\logM \right]
-\frac{M^6}{1769472\pi^6m^4}\,\bigg[ 75\pi^2+26\\
 &\qquad \left.-48\logM +144\logMsq \right] ,\\
     &\tilde\mu^{8-2n}H_2^{(1,1)}(1,1,2,0,2|n+2) \\
 &\quad = -\frac{1}{\epsilon^2}\,\frac{5M^4}{98304\pi^6m^4}
-\frac{1}{\epsilon}\,\frac{M^4}{147456\pi^6m^4}\left[ 1+30\logM
\right]
-\frac{M^4}{1769472\pi^6m^4}\,\bigg[ 87\pi^2\\
 & \qquad \left. +82+48\logM-720\logMsq \right] ,\\
     &\tilde\mu^{8-2n}H_2^{(1,1)}(1,2,1,0,1|n+2) \\
 &\quad= \frac{1}{\epsilon^2}\,\frac{M^4}{49152\pi^6m^2}
-\frac{1}{\epsilon}\,\frac{M^4}{73728 \pi^6m^2}\left[ 1-6\logM
\right]
-\frac{M^4}{884736\pi^6m^2}\,\bigg[ 69\pi^2-26\\
 &\qquad \left.+48\logM-144\logMsq \right] ,\\
     &\tilde\mu^{8-2n}H_2^{(1,1)}(1,2,2,0,1|n+2) \\
 &\quad = \frac{1}{\epsilon^2}\,\frac{11M^4}{98304\pi^6m^4}
-\frac{1}{\epsilon}\,\left[\frac{M^3}{12288\pi^5m^3}-\frac{M^4}{73728\pi^6m^4}\,\left( 5+33\logM \right) \right]\\
 &\qquad +\frac{M^3}{18432\pi^5m^3}\,\left[ 1-\ln8-6\logM \right]
+\frac{M^4}{1769472\pi^6m^4}\bigg[ 190+105\pi^2\\
 &\qquad \left. +480\logM +1584\logMsq \right] ,\\
     & 2 \tilde\mu^{8-2n}H_2^{(1,1)}(1,2,1,0,3|n+4) + \tilde\mu^{8-2n}H_2^{(1,1)}(12202|n+4) \\
 &\quad =-\frac{1}{\epsilon^2}\,\frac{M^6}{786432\pi^8m^4}
+\frac{1}{\epsilon}\,\frac{M^6}{1179648\pi^8m^4}\left[ 1-6\logM
\right] -\frac{M^6}{14155776\pi^8m^4}\,\bigg[ 3\pi^2  \\
 &\qquad\left.+26-48\logM+144\logMsq
\right],\\
     &\tilde\mu^{8-2n}H_2^{(1,1)}(2,1,1,0,3|n+4) \\
 &\quad = \frac{1}{\epsilon^2}\,\frac{M^6}{1572864\pi^8m^4}
-\frac{1}{\epsilon}\,\frac{M^6}{2359296}\left[ 1-6\logM \right]
+\frac{M^6}{28311552\pi^8m^4}\,\bigg[ 27\pi^2\\
 &\quad \left. +26-48\logM+144\logMsq \right] ,\\
     &\tilde\mu^{8-2n}H_2^{(1,1)}(1,1,1,1,1|n+2) \\
 &\quad= \frac{1}{\epsilon}\,\frac{M^5}{6144\pi^5 m^3}
-\frac{M^5}{18432\pi^5 m^3}\left[6\ln(2)-5+12\logM \right]
-\frac{M^6}{12288\pi^4m^4} ,
\end{align*}
   where $\tilde\mu$  is the 't Hooft parameter and we have used
$\tilde\mu=\frac{\mu}{(4\pi)^{1/2}}\,e^{\frac{\gamma_E-1}{2}}$.
   In the above notation the $\widetilde{\mbox{MS}}$ scheme of ChPT corresponds to
subtracting all terms proportional to $\epsilon^{-1}$ and setting
$\mu=m$.

   The corresponding counterterm integrals $H_{CT_1}^{(1,1)}(a,b,c,d,e)$ and
$H_{CT_2}^{(1,1)}(a,b,c,d,e)$ are given by
\begin{align*}
     &\tilde\mu^{8-2n}H_{CT_1}^{(1,1)}(1,1,0,0,1|n) \\
 &\quad = -\frac{1}{\epsilon^2}\frac{3M^4}{1024\pi^4
m^2}-\frac{1}{\epsilon}\frac{M^4}{2048\pi^4
m^2}\left[1+12\logM\right]-\frac{M^4}{4096\pi^4m^2}\left[\pi^2+5+4\logM\right.\\
 &\left.\qquad+24\logMsq\right]\\
 &\quad =\tilde\mu^{8-2n}H_{CT_2}^{(1,1)}(1,1,0,0,1|n),\\
     &\tilde\mu^{8-2n}H_{CT_1}^{(1,1)}(1,1,0,0,2|n+2) \\
 &\quad = \frac{1}{\epsilon^2}\frac{M^6}{24576\pi^6m^2}
-\frac{1}{\epsilon}\frac{M^6}{73728\pi^6m^2}\left[1-6\logM\right]
+\frac{M^6}{884736\pi^6m^2}\bigg[3\pi^2+22\\
 &\left.\qquad  -24\logM+72\logMsq\right]\\
 &\quad =\tilde\mu^{8-2n}H_{CT_2}^{(1,1)}(1,1,0,0,2|n+2),\\
     &\tilde\mu^{8-2n}H_{CT_1}^{(1,1)}(1,2,0,0,2|n+2) \\
 &\quad = \frac{1}{\epsilon^2}\frac{M^4}{16384\pi^6m^2}
+\frac{1}{\epsilon}\frac{M^4}{32768\pi^6m^2}\left[1+4\logM\right]
+\frac{M^4}{196608\pi^6m^2}\bigg[\pi^2+3 +12\logM \\
 &\qquad \left. +24\logMsq \right],\\
     &\tilde\mu^{8-2n}H_{CT_2}^{(1,1)}(1,2,0,0,2|n+2) \\
 &\quad = \frac{1}{\epsilon^2}\frac{M^4}{16384\pi^6m^2}
-\frac{1}{\epsilon}\frac{M^4}{32768\pi^6m^2}\left[1-4\logM\right]
+\frac{M^4}{196608\pi^6m^2}\bigg[\pi^2+9 -12\logM \\
 &\qquad \left. +24\logMsq \right],\\
     &\tilde\mu^{8-2n}H_{CT_1}^{(1,1)}(1,1,1,0,2|n+2) \\
 &\quad = \frac{1}{\epsilon^2}\frac{M^6}{32768\pi^6m^4}
-\frac{1}{\epsilon}\frac{M^6}{589824\pi^6m^4}\left[11-36\logM\right]
+\frac{M^6}{3538944\pi^6m^4}\bigg[9\pi^2+91  \\
 &\qquad \left. -132\logM+216\logMsq \right],\\
     &\tilde\mu^{8-2n}H_{CT_2}^{(1,1)}(1,1,1,0,2|n+2) \\
 &\quad = -\frac{1}{\epsilon^2}\frac{5M^6}{98304\pi^6m^4}
+\frac{1}{\epsilon}\left[\frac{M^5}{12288\pi^5m^3}+\frac{M^6}{589824\pi^6m^4}\left(31-60\logM\right)\right]
\\
 &\qquad -\frac{M^5}{36864\pi^5m^3}\left[5 -6\ln2 -6\logMsq \right]
-\frac{M^6}{3538944\pi^6m^4}\left[15\pi^2+179 -372\logM \right.\\
 &\qquad \left.  +360\logMsq \right],\\
     &\tilde\mu^{8-2n}H_{CT_1}^{(1,1)}(1,2,1,0,1|n+2) \\
 &\quad = \frac{1}{\epsilon^2}\frac{5M^4}{49152\pi^6m^2}
+\frac{1}{\epsilon}\frac{M^4}{98304\pi^6m^2}\left[1+20\logM\right]
+\frac{M^4}{589824\pi^6m^2}\bigg[5\pi^2+27  \\
 &\qquad \left. +12\logM +120\logMsq \right],\\
     &\tilde\mu^{8-2n}H_{CT_2}^{(1,1)}(1,2,1,0,1|n+2) \\
 &\quad = -\frac{1}{\epsilon^2}\frac{M^4}{16384\pi^6m^2}
+\frac{1}{\epsilon}\frac{M^4}{32768\pi^6m^2}\left[1-4\logM\right]
-\frac{M^4}{196608\pi^6m^2}\bigg[\pi^2+9  \\
 &\qquad \left. -12\logM +24\logMsq \right],\\
     &\tilde\mu^{8-2n}H_{CT_1}^{(1,1)}(1,2,2,0,1|n+2) \\
 &\quad = \frac{1}{\epsilon^2}\frac{7M^4}{98304\pi^6m^4}
+\frac{1}{\epsilon}\frac{M^4}{196608\pi^6m^4}\left[1+28\logM\right]
+\frac{M^4}{1179648\pi^6m^4}\bigg[7\pi^2+39  \\
 &\qquad \left. +12\logM +168\logMsq \right],\\
     &\tilde\mu^{8-2n}H_{CT_2}^{(1,1)}(1,2,2,0,1|n+2) \\
 &\quad = \frac{1}{\epsilon^2}\frac{5M^4}{32768\pi^6m^4}
-\frac{1}{\epsilon}\left[\frac{M^3}{12288\pi^5m^3}
+\frac{M^4}{1179648\pi^6m^4}\left(54-360\logM\right) \right]\\
 &\qquad +\frac{M^3}{36864\pi^5m^3}\left[5-6\ln2+6\logM \right] +\frac{M^4}{393216\pi^6m^4}\bigg[5\pi^2+39 -36\logM\\
 &\qquad \left. +120\logMsq \right],\\
     & 2 \tilde\mu^{8-2n}H_{CT_1}^{(1,1)}(1,2,1,0,3|n+4) + \tilde\mu^{8-2n}H_{CT_1}^{(1,1)}(12202|n+4) \\
 &\quad = -\frac{1}{\epsilon^2}\frac{M^6}{786432\pi^8m^4}
+\frac{1}{\epsilon}\frac{M^6}{196608\pi^8m^4}\left[1-6\logM\right]
-\frac{M^6}{28311552\pi^8m^4}\bigg[3\pi^2+22  \\
 &\qquad \left. -24\logM +72\logMsq \right]\\
 &\quad = 2 \tilde\mu^{8-2n}H_{CT_2}^{(1,1)}(1,2,1,0,3|n+4) + \tilde\mu^{8-2n}H_{CT_2}^{(1,1)}(12202|n+4), \\
     &\tilde\mu^{8-2n}H_{CT_1}^{(1,1)}(2,1,1,0,3|n+4) \\
 &\quad = -\frac{1}{\epsilon^2}\frac{M^6}{4718592\pi^8m^4}
+\frac{1}{\epsilon}\frac{M^6}{28311552\pi^8m^4}\left[5-12\logM\right]
-\frac{M^6}{169869312\pi^8m^4}\bigg[3\pi^2  \\
 &\qquad \left. +37 -60\logM +72\logMsq \right],\\
     &\tilde\mu^{8-2n}H_{CT_2}^{(1,1)}(2,1,1,0,3|n+4) \\
 &\quad = \frac{1}{\epsilon^2}\frac{7M^6}{4718592\pi^8m^4}
-\frac{1}{\epsilon}\frac{M^6}{28311552\pi^8m^4}\left[17-84\logM\right]
+\frac{M^6}{169869312\pi^8m^4}\bigg[21\pi^2  \\
 &\qquad \left. +169 -204\logM +504\logMsq \right],\\
     &\tilde\mu^{8-2n}H_{CT_1}^{(1,1)}(1,1,1,1,1|n+2) \\
 &\quad = \frac{1}{\epsilon^2}\left[\frac{M^5}{12288\pi^5m^3}+\frac{M^6}{36864\pi^6m^4}
\right]
-\frac{1}{\epsilon}\frac{M^6}{884736\pi^6m^4}\left[11-48\logM\right]\\
 &\quad -\frac{M^5}{36864\pi^5m^3}\left[5 -6\ln(2) -6\logM \right]\\
 &\quad = \tilde\mu^{8-2n}H_{CT_2}^{(1,1)}(1,1,1,1,1|n+2).
\end{align*}

   The results for the products of one-loop integrals read
\begin{align*}
     &\tilde\mu^{8-2n}H_2^{(1,1)}(1,1,0,0,0|n) \\
 &\quad = -\frac{1}{\epsilon^2}\,\frac{M^4}{256\pi^4}
-\frac{1}{\epsilon}\,\frac{M^4}{64\pi^4}\logM
-\frac{M^4}{1536\pi^4}\,\left[ \pi^2+6+48\logMsq \right],\\
     &\tilde\mu^{8-2n}H_2^{(1,1)}(1,2,0,0,0|n) \\
 &\quad = -\frac{1}{\epsilon^2}\,\frac{M^2}{256\pi^4}
-\frac{1}{\epsilon}\,\frac{M^2}{256\pi^4}\left[ 1+\logM \right]
-\frac{M^2}{1536\pi^4}\,\left[ \pi^2+6+24\logM+48\logMsq \right] ,\\
     &\tilde\mu^{8-2n}H_2^{(1,1)}(1,1,1,0,0|n) \\
 &\quad= -\frac{1}{\epsilon^2}\,\frac{M^4}{512\pi^4m^2}
-\frac{1}{\epsilon}\,\left[ \frac{M^3}{256\pi^3
m}+\frac{M^4}{512\pi^4m^2}\left(1+4\logM\right) \right]
+\frac{M^3}{256\pi^3 m}\,\bigg[1-2\ln(2)  \\
 &\qquad \left. -4\logM  \right] -\frac{M^4}{3072\pi^4m^2}\,\left[\pi^2+6+24\logM+48\logMsq\right],\\
     &\tilde\mu^{8-2n}H_2^{(1,1)}(1,2,0,1,0|n) \\
 &\quad= -\frac{1}{\epsilon^2}\,\frac{M^2}{512\pi^4m^2}
-\frac{1}{\epsilon}\,\left[\frac{M}{512\pi^3m}+\frac{M^2}{256\pi^4m^2}\left(
1+2\logM \right)\right] \! -\frac{M}{512\pi^3m}\left[ 1+2\ln(2) \right.\\
 & \qquad \left. +4\logM \right] -\frac{M^2}{3072\pi^4m^2}\,\left[ \pi^2 +12 +48\logM+48\logMsq\right] ,\\
     &\tilde\mu^{8-2n}H_2^{(1,1)}(2,1,0,2,0|n+2) \\
 &\quad= \frac{1}{\epsilon^2}\,\frac{M^4}{8192\pi^6m^2}
+\frac{1}{\epsilon}\,\frac{M^4}{2048\pi^6m^2}\logM
+\frac{M^4}{49152\pi^6m^2}\,\left[ \pi^2+6+48\logMsq \right] ,\\
     &\tilde\mu^{8-2n}H_2^{(1,1)}(1,1,1,1,0|n) \\
 &\quad= -\frac{M^2}{256\pi^2m^2} ,\\
     &\tilde\mu^{8-2n}H_2^{(1,1)}(1,1,1,1,0|n+2) \\
 &\quad= -\frac{M^6}{9216\pi^4m^2}.
\end{align*}
   The integral $H_2^{(1,1)}(1,0,0,0,1|n)$ can be written as
$H_2^{(1,1)}(1,0,0,1,0|n)$ by the substitution $k_2\mapsto
k_2+k_1$, and
$$
   H_2^{(1,1)}(1,0,0,1,0|n)=H_1^{(1)}(1,0|n)H_1^{(1)}(0,1|n)=0,
$$
since the infrared singular part $H_1^{(1)}(01|n)$ of
$H_1(01|n)=-i I_N$ vanishes.

   The counterterm integrals corresponding to the products of one-loop integrals
read
\begin{align*}
     &\tilde\mu^{8-2n}H_{CT_1}^{(1,1)}(1,1,0,0,0|n) \\
 &\quad = -\frac{1}{\epsilon^2}\frac{M^4}{256\pi^4}
-\frac{1}{\epsilon}\frac{M^4}{128\pi^4}\logM
-\frac{M^4}{3072\pi^4}\left[\pi^2+6+24\logMsq\right]\\
 &\quad =\tilde\mu^{8-2n}H_{CT_2}^{(1,1)}(1,1,0,0,0|n),\\
     &\tilde\mu^{8-2n}H_{CT_1}^{(1,1)}(1,2,0,0,0|n) \\
 &\quad = -\frac{1}{\epsilon^2}\frac{M^2}{256\pi^4}
-\frac{1}{\epsilon}\frac{M^2}{256\pi^4}\left[1+2\logM\right]
-\frac{M^2}{3072\pi^4}\left[\pi^2+6+24\logM+24\logMsq\right],\\
     &\tilde\mu^{8-2n}H_{CT_2}^{(1,1)}(1,2,0,0,0|n) \\
 &\quad = -\frac{1}{\epsilon^2}\frac{M^2}{256\pi^4}
-\frac{1}{\epsilon}\frac{M^2}{128\pi^4}\logM
-\frac{M^2}{3072\pi^4}\left[\pi^2+6+24\logMsq\right],\\
     &\tilde\mu^{8-2n}H_{CT_1}^{(1,1)}(1,1,1,0,0|n) \\
 &\quad = -\frac{1}{\epsilon^2}\frac{M^4}{512\pi^4m^2}
-\frac{1}{\epsilon}\frac{M^4}{256\pi^4m^2}\logM
-\frac{M^4}{6144\pi^2m^2}\left[\pi^2+6+24\logMsq\right],\\
     &\tilde\mu^{8-2n}H_{CT_2}^{(1,1)}(1,1,1,0,0|n) \\
 &\quad = -\frac{1}{\epsilon^2}\frac{M^4}{512\pi^4m^2}
-\frac{1}{\epsilon}\left[\frac{M^3}{256\pi^3m}+\frac{M^4}{512\pi^4m^2}\left(1+2\logM\right)\right]
+\frac{M^3}{256\pi^3m}\bigg[1-2\ln(2) \\
 &\qquad \left.+2\logM\right] -\frac{M^4}{6144\pi^2m^2}\left[\pi^2+6+24\logM+24\logMsq\right],\\
     &\tilde\mu^{8-2n}H_{CT_1}^{(1,1)}(1,2,0,1,0|n) \\
 &\quad = -\frac{1}{\epsilon^2}\frac{M^2}{512\pi^4m^2}
-\frac{1}{\epsilon}\left[\frac{M}{512\pi^3m}+\frac{M^2}{256\pi^4m^2}\left(1+\logM\right)\right]
-\frac{M}{512\pi^3m}\bigg[1+2\ln(2)\\
 &\qquad \left. +2\logM\right]  -\frac{M^2}{6144\pi^4m^2}\left[ \pi^2 +18 +48\logM+24\logMsq \right],\\
     &\tilde\mu^{8-2n}H_{CT_2}^{(1,1)}(1,2,0,1,0|n) \\
 &\quad = -\frac{1}{\epsilon^2}\frac{M^2}{512\pi^4m^2}
-\frac{1}{\epsilon}\frac{M^2}{256\pi^4m^2}\logM
-\frac{M^2}{6144\pi^4m^2}\left[\pi^2+6+24\logMsq\right],\\
     &\tilde\mu^{8-2n}H_{CT_1}^{(1,1)}(2,1,0,2,0|n+2) \\
 &\quad = \frac{1}{\epsilon^2}\frac{M^4}{8192\pi^6m^2}
+\frac{1}{\epsilon}\frac{M^4}{4096\pi^6m^2}\logM
+\frac{M^4}{98304\pi^6m^2}\left[\pi^2+6+24\logMsq\right]\\
 &\quad = \tilde\mu^{8-2n}H_{CT_2}^{(1,1)}(2,1,0,2,0|n+2), \\
     &\tilde\mu^{8-2n}H_{CT_1}^{(1,1)}(1,1,1,1,0|n) \\
 &\quad = 0\\
 &\quad = \tilde\mu^{8-2n}H_{CT_2}^{(1,1)}(1,1,1,1,0|n), \\
     &\tilde\mu^{8-2n}H_{CT_1}^{(1,1)}(1,1,1,1,0|n+2) \\
 &\quad = 0\\
 &\quad = \tilde\mu^{8-2n}H_{CT_2}^{(1,1)}(1,1,1,1,0|n+2). \\
\end{align*}

\end{document}